\documentclass[11pt,a4paper]{article}

\usepackage{amsmath}

\usepackage{geometry}
\usepackage{graphicx}
 \usepackage{epsf,epsfig,psfrag,graphics,color}
 
 \usepackage{amsmath}
 \usepackage{amssymb}
 \usepackage{appendix}
 \usepackage{multirow}
 \usepackage{geometry}
 \usepackage{float}

\def\kp{_{K\pi}}

\def\mo{\stackrel{\circ}{M}}

\begin{document}
\thispagestyle{empty}

\begin{flushright}
LPT-ORSAY/10-60\\
\end{flushright}

\vspace{\baselineskip}

\begin{center}
\vspace{3\baselineskip}
\textbf{\Large 
Chiral dynamics with strange quarks\\[0.5em] 
in the light of recent lattice simulations
}\\

\vspace{3\baselineskip}
{\sc V\'eronique Bernard$^{(a)}$, S\'ebastien Descotes-Genon$^{(b)}$ and Guillaume Toucas$^{(b)}$}\\

\vspace{0.7cm}
{\sl $(a)$ Institut de Physique 
Nucl\'eaire, CNRS/Univ. Paris-Sud 11 (UMR 8608),\\
91406 Orsay Cedex, France\\

$(b)$ Laboratoire de Physique 
Th\'eorique, CNRS/Univ. Paris-Sud 11 (UMR 8627),\\
\emph{91405 Orsay Cedex, France}}
\vspace{3\baselineskip}

\vspace*{0.5cm}
\textbf{Abstract}\\
\vspace{1\baselineskip}
\parbox{0.9\textwidth}{
Several lattice collaborations performing simulations with 2+1 light dynamical quarks
have experienced difficulties in fitting their data with standard $N_f=3$ 
chiral expansions at next-to-leading order, yielding low values of the quark 
condensate and/or the 
decay constant in the $N_f=3$ chiral limit. A reordering of these expansions
seems required to analyse these data in a consistent way. We 
discuss such a reordering, known as Resummed Chiral Perturbation 
Theory, in the case of pseudoscalar masses and decay constants, pion and kaon 
electromagnetic form factors and $K_{\ell 3}$ form factors. We show that it 
provides a good fit of the recent results of two lattice collaborations (PACS-CS and 
RBC/UKQCD). We describe the emerging picture for the pattern of chiral 
symmetry breaking, marked by a strong dependence of the observables on the 
strange quark mass and thus a significant difference between chiral symmetry 
breaking in the $N_f=2$ and $N_f=3$ chiral limits. We discuss 
the consequences for the ratio of decay constants $F_K/F_\pi$ and the 
$K_{\ell 3}$ form factor at vanishing momentum transfer.
} 
\end{center}

\clearpage


\newpage
\setcounter{page}{1}

\section{Introduction}

Recent improvements in lattice simulations have paved the way for 
unquenched simulations with three light dynamical (or sea) quarks. Even though 
these simulations are still at an early stage, their results are already 
very instructive, especially when one is interested in the low-energy dynamics 
of strong interactions. Indeed, there is a deep connection between these 
first-principle simulations and the effective description of the light hadronic
degrees of freedom, i.e., Chiral Perturbation Theory ($\chi$PT)~\cite{chpt-su2,
chpt-su3}. The ability for the lattice to investigate the quark-mass dependence
of hadronic observables opens  the possibility to determine some poorly known 
$\chi$PT low-energy constants. Conversely, the chiral expansions derived within
$\chi$PT can be used to extrapolate or interpolate lattice data down to the 
physical light quark masses. 

It turns out that several collaborations performing simulations
with 2+1 dynamical quarks reported difficulties when fitting their data with 
$N_f=3$ next-to-leading-order (NLO) chiral expansions
for pseudoscalar masses, decay constants, and $K_{\ell 3}$ form
factors~\cite{
Aoki:2008sm,Allton:2008pn,Boyle:2007qe,Boyle:2010bh}, forcing some of the
collaborations to rely on $N_f=2$ chiral expansions only (for instance using
on Heavy-Kaon $\chi$PT to perform their extrapolations~\cite{Flynn:2008tg}). 
Moreover,
some chiral order parameters (in particular the quark condensate and the decay 
constant) were seen to decrease significantly when one moved from the
$N_f=2$ 
chiral limit where $m_u,m_d\to 0$ but $m_s$ is kept at its physical 
value~\cite{chpt-su2} to the $N_f=3$ chiral limit  
where $m_u,m_d, m_s \to 0$~\cite{chpt-su3}. 
Defining the chiral order parameters in these two chiral limits (denoted
$\lim_{N_f}$ for simplicity):
\begin{equation}
\Sigma(N_f)=-{\rm{lim}}_{N_f} \langle 0 |\bar{u}u| 0\rangle\,, \qquad
F^2(N_f)={\rm{lim}}_{N_f} F^2_\pi\,,
\end{equation}
{\rm PACS-CS}~\cite{Aoki:2008sm} and {\rm MILC}~\cite{Bernard:2007ps,
Bazavov:2009tw} quote for instance:
\begin{eqnarray}
{\rm PACS-CS}&:& \frac{F(2)}{F(3)}=1.089\pm 0.045\,, \qquad\qquad\qquad \frac{\Sigma(2)}{\Sigma(3)}=1.245\pm 0.010\,,\\
{\rm MILC} &:& \frac{F(2)}{F(3)}=1.15\pm 0.05\left(\begin{array}{c}+0.10\\-0.03\end{array}\right)\,, \qquad \frac{\Sigma(2)}{\Sigma(3)}=1.52 \pm 0.17 \left(\begin{array}{c}+0.38\\-0.15\end{array}\right)\,.
\end{eqnarray}

\noindent Assuming that there are no intrinsic problems with the lattice data,
these results can be interpreted as the fact that the $N_f=3$ chiral series do not 
converge quickly, because their leading-order term in the chiral counting is 
not numerically dominant and competes with (formally) higher-order contributions. 
Calculations up to NNLO are clearly useful to settle the issue~\cite{Bijnens:2006zp,Bijnens:2006mk,Bijnens:2006jv,Bijnens:2006ca}. An exploratory
study of the RBC/UKQCD data~\cite{Allton:2008pn,Boyle:2007qe} on $K_{\ell 3}$ form
factors and $F_K/F_\pi$ (the ratio of the kaon and pion decay constants) showed that 
NNLO terms are not negligible though of the expected size, the NLO terms of some 
observables being in this particular study anomalously small for some known reasons.  
However, this study did not consider the RBC/UKQCD result for the decay constant $F_\pi$ itself, which does not exhibit the quark-mass dependence predicted by chiral perturbation theory~\cite{Allton:2008pn,Mawhinney:2009jy}. 

Quite interestingly, a similar pattern with suppressed $N_f=3$ chiral order parameters
seems to emerge from the experimental 
data on $\pi\pi$ and $\pi K$ scattering, indicating a good convergence of 
$N_f=2$ chiral series for pion observables~\cite{pipiampl,Ananthanarayan:2000ht,
Colangelo:2001sp,pipidata,Bijnens:2004eu,Batley:2007zz,:2009nv,Colangelo:2008sm}, 
but difficulties in $N_f=3$ chiral series, even 
once NNLO terms are included and modeled through resonance 
saturation~\cite{roypika,Bijnens:2004bu}. As recalled in 
refs.~\cite{Bijnens:2006zp,DescotesGenon:2007bs,Bijnens:2009hy}, the situation 
seems to be well under control in the $\pi\pi$ system, but the fits yield  
sometimes  contradictions once $\pi K$ data is considered.
It was advocated in refs.~\cite{paramag,vaclecs} that such a situation could 
indeed occur in the low-energy dynamics of QCD
and it was  highlighted there 
that the strange quark may play a very special role, due to its light mass 
of order $O(\Lambda_{QCD})$. 
Significant vacuum fluctuations of $s\bar{s}$ pairs may lead to 
instabilities in the convergence of
$\chi$PT expansions, where instabilities are defined as a numerical competition
between the terms considered as leading and next-to-leading in the
chiral counting. This effect would be related to a large violation
of the Zweig rule in the scalar sector, indicated by values
of the $O(p^4)$ LECs $L_4$ and $L_6$ significantly different from
specific (``critical'') values~\cite{bachirl6,sumrulel6} (see also refs.~\cite{mei02,ber07} for reviews of the subject and a discussion of the baryon sector).

A pessimistic way of considering the problem would consist in dismissing
the whole $\chi$PT as soon as problems of convergence arise. 
A less restrictive point of view was adopted in ref.~\cite{resum}, assuming that:
\begin{itemize}
\item only some (``good'') observables have convergent expansions, when expressed in terms of the couplings arising in the chiral Lagrangian.
\item a series is considered as convergent when the sum of LO and NLO terms is large compared to the remaining part of the series.
\item the resulting formulae must be treated analytically, without neglecting higher-order corrections
when reexpressing low-energy constants in terms of observables.
\end{itemize}
This framework was coined Resummed Chiral Perturbation Theory (Re$\chi$PT) for reasons that will be recalled later in the present article. In particular, a specific prescription for the unitarity pieces, described in sec.~\ref{sec:emff}, will be added to this set of assumptions, defining precisely how we choose to split the chiral series of the chosen observables between leading, next-to-leading and higher orders in our framework.

The recent $2+1$ dynamical simulations provide new and 
relevant  information on  the impact of $s\bar{s}$ fluctuations related to the presence of strange 
quarks in the sea. Conversely,
Re$\chi$PT can provide a more appropriate treatment of chiral 
extrapolations if the hints of suppressed $N_f=3$ quark condensate and decay 
constants are confirmed. In principle, this analysis would require lattice 
data performed with several $u,d,s$ quark masses (whose renormalized values 
are known) and transfer momenta (in the case of form factors, scattering amplitudes), but where the 
continuum and infinite-volume limit have already been performed ($a\to 0, L\to\infty$). 
Unfortunately, such data sets are not (yet) available. Some collaborations 
[e.g., MILC~\cite{Bernard:2007ps,Bazavov:2009tw}] provide numbers directly in 
the physical limit, performing the chiral extrapolation at the same time as 
the continuum limit. This prevents one from testing different alternatives 
concerning chiral extrapolation, even though the results sometimes 
contradict the usual assumptions of $\chi$PT (for instance concerning the size of the 
quark condensate and the decay constant). Others 
[BMW~\cite{Durr:2010hr,Durr:2008zz}, ETMC~\cite{Baron:2010bv}, 
TWQCD-JLQCD~\cite{Noaki:2009sk}] do not provide the decay constants and 
the renormalized quark masses mandatory for such a study. Finally, some 
collaborations [RBC/UKQCD~\cite{Allton:2008pn,Boyle:2007qe,Boyle:2010bh}, 
PACS-CS~\cite{Aoki:2008sm}] have performed their analysis only at one 
particular lattice spacing and/or one particular volume, without estimating 
the systematics associated with the continuum and infinite-volume limits fully.

In view of this situation, we will restrict ourselves to the studies made 
by RBC/UKQCD and PACS-CS. The fact that only statistical errors are quoted in 
both cases prevents us from using a full-fledged statistical 
treatment~\cite{DescotesGenon:2007bs}, but we hope that our limited study 
will provide some incentive for lattice collaborations to apply the same 
framework to their data when performing chiral extrapolations.
In Sec.~\ref{sec:basics} we provide the basics of our procedure with the 
illustration of 
the electromagnetic form factor of the pion, and we recall the results obtained for the
masses and decay constants. In Secs.~\ref{sec:kaon} and \ref{sec:Kl3}, 
we apply the same formalism to the electromagnetic form factor of the kaon
and the $K_{\ell 3}$ form factors.
In Sec.~\ref{sec:lattice}, we consider the same observables in the case of 
lattice simulations for non-physical values of the quark masses. 
In Sec.~\ref{sec:fit}, we fit  our chiral expansions to different sets of 
lattice data, and analyse the emerging pattern of $N_f=3$ chiral symmetry 
breaking, before concluding. An appendix summarises the main features of the 
results obtained by the two lattice collaborations which are analysed in this 
article.

\section{General arguments and the electromagnetic form factor of the pion}
\label{sec:basics}
\subsection{Expansion of "good" observables} \label{sec:good}

As explained in greater details in 
refs.~\cite{ordfluc,resum,DescotesGenon:2007bs}, if one expects a numerical 
competition between LO and NLO chiral series of quantities of interest, 
one cannot perform expansions on arbitrary functions of such quantities. 
Let us assume for instance that an observable $A$ has the following 
chiral expansion:
\begin{equation}
A=A_{LO}+A_{NLO}+A\delta A
\end{equation}
The statement of a good convergence of the series correspond to $\delta A$ 
much smaller than 1, but it does not involve necessarily that $A_{LO}$ is 
dominant numerically with respect to the rest of the series.

One has  the formal chiral expansion for any observable $F=f(A)$:
\begin{eqnarray}
F&=&F_{LO}+F_{NLO}+F\delta F\,,\\
F_{LO}&=&f(A_{LO}) \qquad\qquad\qquad\qquad F_{NLO}=A_{NLO}f'(A_{LO})\,,\\
\delta F&=&1-\frac{f(A_{LO})}{f(A)}-\frac{f'(A_{LO})}{f(A)}[A-A_{LO}-A\delta A]\,.
\end{eqnarray}
Depending on the value of $X_A=A_{LO}/A$, i.e., the saturation of the chiral 
expansion of $A$ by its first term, the chiral series of $F$ may or may not 
converge well. One has in particular the two limiting behaviours:
\begin{eqnarray}
X_A\to 1 &\qquad & \delta F \to - \frac{f'(A)}{f(A)}A\delta A\,,\\
X_A\to 0 &\qquad & \delta F \to 1-\frac{f(0)}{f(A)}-\frac{f'(0)}{f(A)}A+\frac{f'(0)}{f(A)}A\delta A\,.
\end{eqnarray}
In the first case, a bound on $\delta A$ implies a bound on $\delta F$, 
meaning that $F$ converges well provided that $A$ does. But in the second case, the size of $\delta F$ is by no means driven by that of $\delta A$. As an 
illustration, if we take the "observables"
$B=1/A$ and $C=\sqrt{A}$, we obtain:
\begin{eqnarray}
\delta B &=& \frac{(1-X_A)^2}{X_A^2}-\frac{\delta A}{X_A^2}\,,\\
\delta C &=& 1-\frac{1}{2\sqrt{X_A}}-\frac{1}{2}\sqrt{X_A}-\frac{1}{2\sqrt{X_A}}\delta A\,.
\end{eqnarray}
Even if we set $\delta A=0$, we  would need $X_A=A_{LO}/A$ above $41\%$ to ensure 
that $|\delta B|<10\%$, and $X_A$ above 76\% to ensure that $|\delta C|<10\%$.

Therefore, if the chiral expansions of some of the observables considered are not 
saturated by their leading-order term, we cannot take 
an arbitrary function of these observables, consider its chiral expansion, and 
assume that it will converge.
In such a situation, one has therefore to select the right set of observables 
for which converging series can be written. Actually, this statement of 
convergence is equivalent to state that the generating functional of QCD is 
well reproduced at low energies by $\chi$PT. The quantities 
that are assumed to have convergent chiral series from the starting point of 
chiral perturbation theory are the QCD correlators of axial and vector 
currents as well as pseudoscalar and scalar densities. In addition, the 
convergence is expected to be good only away from the singularities 
(poles, cuts \ldots) corresponding to resonances and channel openings. Observables 
involving pseudoscalar mesons as external states will be obtained by applying 
the LSZ reduction formula on correlators involving axial currents, which will 
yields additional factors of the pseudoscalar decay constants. 
One can easily state  some of the observables associated to each
$O(p^4)$ LEC that are simple to determine on lattice simulations and/or to extract from experiment:
\begin{itemize}
\item $L_1,L_2,L_3$: $K_{\ell 4}$ form factors ($F_\pi^2 F_K f,F_\pi^2 F_K g,F_\pi^2 F_K h$),
\item $L_4,L_5$: $\pi,K$ decay constants ($F_\pi^2,F_K^2$),
\item $L_6,L_8$: $\pi,K$ masses ($F_\pi^2M_\pi^2,F_K^2M_K^2$),
\item $L_7$: $\eta$ mass and decay constant ($F_\eta^2, F_\eta^2M_\eta^2$),
\item $L_9$: Pion electromagnetic form factor ($F_\pi^2 F^\pi_V$),
\item $L_{10}$: $\pi\to e\nu\gamma$ form factor ($F_\pi A_{VAP}$) and
$\tau$ spectral functions ($VV-AA$).
\end{itemize}

We will define the expansion of a ``good'' observable through
the following procedure:
\begin{itemize}
\item We take the chiral expansion of the observable in terms of the
couplings of the effective Lagrangian (low-energy constants, or LECs: $B_0,F_0,L_i$\ldots).
\item We replace the pseudoscalar masses at leading order by the physical ones only when physical arguments indicate that the convergence of the series will be improved. In practice we perform this replacement so that the nonanalytic structures imposed by unitarity are located at the physical
poles, thresholds\ldots
\item We keep track of the remainders (collecting NNLO contributions and higher orders) explicitly and 
treat the resulting chiral expansions as algebraic identities, 
without performing further expansions.
\end{itemize}

Following this procedure (and taking the isospin limit $m_u=m_d=m$), 
observables can be expressed in terms of LO quantities:
\begin{equation}\label{eq:LOLECs}
X(3)=\frac{2m\Sigma(3)}{F_\pi^2M_\pi^2}\,,\qquad Z(3)=\frac{F^2(3)}{F_\pi^2}\,,\qquad r=\frac{m_s}{m}\,,
\end{equation}
as well as NLO LECs and remainders. The first two quantities in 
eq.~(\ref{eq:LOLECs}) are of particular relevance, since they express 
two main order parameters of $N_f=3$ chiral symmetry breaking, the quark 
condensate and the pseudoscalar decay constant, in physical units. They also 
assess the saturation of the chiral expansion of $F_\pi^2M_\pi^2$ and 
$F_\pi^2$ by their leading order. The third quantity 
measures the relative size of the quark masses in a framework where the 
strange quark is supposed to play a peculiar role in the chiral structure of 
QCD vacuum. In the following, we will also use the quantity
$Y(3)$:
\begin{equation}
Y(3)=\frac{2mB_0}{M_\pi^2}=\frac{X(3)}{Z(3)}\,,
\label{eq:y3}
\end{equation}
which assesses the saturation of the chiral expansion of the pion mass by its leading order.
Up to now, we have considered the masses and decay
constants of Goldstone bosons~\cite{vaclecs,sumrulel6,ordfluc} and observables
derived from $\pi\pi$ scattering~\cite{ordfluc,resum} in this framework. 

Exploiting the fact that some quantities describing the dynamics of pseudoscalar mesons are well measured and inverting
the relationships between these observables and LECs, we can express the $O(p^4)$ LECs in terms of 
\begin{itemize}
\item Masses, decay constants, form factors\ldots
\item The three leading-order parameters described in eq.~(\ref{eq:LOLECs})
\item The remainders associated to each observable, assumed to be small (convergence).
\end{itemize}
These expressions can be exploited in the chiral expansions of other "good" 
observables, in order to express the latter quantities in terms of LO 
quantities and  remainders only. The comparison with experimental 
information should then provide more  information on the pattern of chiral symmetry breaking in the $N_f=3$ chiral 
limit.

As shown in refs.~\cite{ordfluc,resum,DescotesGenon:2007bs}, this procedure 
applied to masses and decay constants allows one to resum higher-order 
contributions in chiral series from $L_4$ and $L_6$ low-energy constants, 
which encode the effect of $s\bar{s}$ pairs on the structure of the chiral 
vacuum. It may induce a significant $m_s$~-dependence in the pattern of chiral 
symmetry breaking and can generate a numerical competition between LO and 
NLO in $N_f=3$ chiral series. This feature is related to the name of 
Resummed Chiral Perturbation Theory chosen to describe this 
particular treatment of chiral expansions. This framework is compatible with the usual 
treatment of chiral series in the limit where the latter are saturated by their LO term, but it allows for a consistent treatment of 
the series even if there is a significant competition of LO and NLO 
contributions for some of the observables.

\subsection{Electromagnetic form factor}\label{sec:emff}

We will illustrate our procedure with the example of the electromagnetic pion form factor
\begin{equation}
\langle \pi^+|j_\mu|\pi^+\rangle = (p+p')^\mu F^\pi_V(t)\,,
\end{equation}
where the electromagnetic current is $j_\mu=V_\mu^3+V_\mu^8/\sqrt{3}$, 
$p$ ($p'$) is the momentum of the incoming (outgoing) pion,
and $t=(p'-p)^2$.
As explained in Sec.~\ref{sec:good}, we obtain this form factor in $\chi$PT from
$\langle (A^\mu_{\pi^+})^\dag A^\nu_{\pi^+} j^\mu\rangle$, leading
to the product $F_\pi^2 F^\pi_V$ through the LSZ reduction formula
(the pion decay constant $F_\pi$
stems from the wave-function renormalisation). In
the case of the electromagnetic form factor, ``good'' observables
are thus obtained from $F_\pi^2 F^\pi_V$ at low energies away from  
singularities, i.e., the right-hand cuts, starting from $t\geq 4 M_\pi^2$.

We obtain the following bare expansion, in agreement with Refs.~\cite{ffactors,ffactors-pion,ffactors-em-p6}:
\begin{eqnarray}\label{eq:pivect-bare}
F_\pi^2 F^\pi_V(t)
 &=& F_\pi^2Z(3) + 
       M_\pi^2 Y(3)
          [8(r+2)L_4^r+8L_5^r]
          \nonumber
    -\frac{1}{32\pi^2} M_\pi^2 Y(3)
          \Bigg[4\log\frac{\mo_\pi^2}{\mu^2}
                +(r+1)\log\frac{\mo_K^2}{\mu^2}\Bigg]\\
\nonumber
&&\quad
+t
   \Bigg[2L_9^r-\frac{1}{32\pi^2}
                 \left[\frac{1}{3}\log\frac{\mo_\pi^2}{\mu^2}
                      +\frac{1}{6}\log\frac{\mo_K^2}{\mu^2}
                      +\frac{1}{6}\right]\Bigg]\\
&&\quad
+\frac{1}{6}[t-4M_\pi^2Y(3)]\bar{J}_{\pi\pi}(t)
        +\frac{1}{12}[t-2(r+1)M_\pi^2 Y(3)]\bar{J}_{KK}(t)
 + \Re F_\pi^2F^\pi_V(t) \,,
\end{eqnarray}
with the nonanalytic pieces from the two-meson channels  
encoded in the $\bar{J}$ function~\cite{chpt-su3}.
$\mo_P$ denotes the leading order of the chiral expansion of $O(p^2)$:
\begin{equation}
\mo_\pi^2=M_\pi^2 Y(3) \qquad \mo_K^2=M_\pi^2\frac{r+1}{2} Y(3)
 \qquad \mo_\eta^2=M_\pi^2\frac{1}{3}(2r+1) Y(3)\,,
\end{equation}
and $Y(3)$ is the ratio defined in eq.(\ref{eq:y3}). We have added 
$\Re F_\pi^2F^\pi_V(t)$, a polynomial function of $t$ collecting
remainders:
\begin{equation}
 \Re F_\pi^2F^\pi_V(t)
  = (\Re F_\pi^2F^\pi_V)_0 + \frac{t}{F_\pi^2} (\Re F_\pi^2F^\pi_V)_1 + O(t^2)\,,\end{equation}
with  
$(\Re F_\pi^2F^\pi_V)_0 = O(m_q^2)$ and $(\Re F_\pi^2F^\pi_V)_1 = O(m_q)$.

At next-to-leading order in the chiral expansion,
the nonanalytic dependence on quark masses and momenta arises through
the unitarity  function $\bar{J}_{PQ}$.
Following our prescription, we compute
the functions $\bar{J}$ (and $\bar{\bar{J}}=\bar{J}-s\bar{J}'(0)$) with
the physical values of $M_\pi^2, M_K^2, M_\eta^2$, rather than their leading-order expansion, i.e. we define the chiral expansion in Re$\chi$PT as 
eq.~(\ref{eq:pivect-bare}) with:
\begin{equation}\label{eq:JbarPP}
\bar{J}_{PP}(t)=\frac{s}{16\pi^2}
      \int_{4M_P^2}^\infty \frac{dx}{x(x-s)} \sqrt{1-\frac{4M_P^2}{x}}
  =\frac{1}{16\pi^2}
    \left[\sigma \log\frac{\sigma-1}{\sigma+1} + 2\right]
\qquad \sigma=\sqrt{1-\frac{4M_P^2}{t}}\,,
\end{equation}
so that our expansion of the form factor eq.~(\ref{eq:pivect-bare}) features a unitarity cut from the two-pion channel starting at $t=4M_\pi^2$ (id. for the two-kaon channel).
Indeed, from general arguments of unitarity, we know that the higher-order 
corrections will shift the start of the right-hand cut from $4\mo_P^2$ to 
$4M_P^2$. Unfortunately, unitarity does not provide us more information on the structure of the cut (and in particular the coefficient multiplying the $\bar{J}$ function) due to the perturbative nature of the chiral expansion.

When needed, we will obtain the Re$\chi$PT expansion of other observables by 
performing the same replacement for the functions $\bar{J}$ and 
$\bar{\bar{J}}$ occurring in the definition of the loop integrals 
$K_{PQ}$, $L_{PQ}$ and $M^r_{PQ}$ in ref.~\cite{chpt-su3}.
However, we do \emph{not} perform any
further replacement neither in the unitary functions nor in the rest of the 
expressions: for instance, we have not modified the functions multiplying the 
$\bar{J}$ functions, nor the chiral logarithms coming from the tadpole terms 
in eq.~(\ref{eq:pivect-bare}), since we have no way of determining if the latter 
modifications would improve or spoil the convergence of the 
series~\footnote{This procedure is slightly different from the approach taken 
in Ref.~\cite{DescotesGenon:2007bs}, where this substitution was performed 
everywhere in the unitarity functions $J,K,L,M$ and in the tadpole logarithms. 
It turns out that the difference is usually very small: the unitarity 
functions yield only a small contribution below the first threshold, and there 
is only a logarithmic difference in the case of the tadpole.}. Adopting a dispersive point of view, 
we can say that the position of the cuts are imposed by unitarity, but not 
the value of the induced imaginary parts at low energies and that 
of the subtraction constants (polynomials).

One checks easily that the NLO chiral expansion of the electromagnetic 
form factor in ref.~\cite{ffactors} can be recovered:
\begin{equation}
F^\pi_V(t)= 1+2H_{\pi\pi}(t)+H_{KK}(t)+O(p^4)\,,
\label{eq:fpiv}
\end{equation}
with 
\begin{equation}
H_{PP}(t)=\frac{1}{F_0^2}
   \left[
  \frac{1}{12}\left(t-4M_P^2\right)\bar{J}_{PP}(t)
                     -\frac{t}{6}\frac{1}{32\pi^2}\left(\log\frac{M_P^2}{\mu^2}+1\right)+\frac{t}{288\pi^2}\right]
          +\frac{2t}{3F_0^2}L_9^r\,.
\label{eq:hpp}
\end{equation}
In the case where $F_0$ is small compared to $F_\pi$, as hinted at by lattice simulations and NNLO fits of chiral series~\cite{Bijnens:2009hy,Bernard:2009ds},
$F_\pi^2 F^\pi_V(t)$ is expected to exhibit a better convergence than $F^\pi_V(t)$ in our framework according to Eqs.~(\ref{eq:fpiv})-(\ref{eq:hpp}). Similar expressions hold for other observables: good observables will generally come multiplied by powers of physical pseudoscalar decay constants (one for each external pseudoscalar meson involved).

\subsection{Pion electromagnetic square radius}

The electromagnetic square radius of the pion is the low-energy observable associated with $F^\pi_V$:
\begin{equation}
F_\pi^2 \langle r^2\rangle^\pi_V=
  6F_\pi^2 \frac{dF^\pi_V}{dt}(0)\,.
  \end{equation}
Following the previous discussion of the form factor, the product of $F_\pi^2$ and $\langle r^2\rangle^\pi_V$ is the quantity expected to exhibit a good convergence in our framework.
eq.~(\ref{eq:pivect-bare}) yields the corresponding expansion of $\langle r^2\rangle^\pi_V$:
\begin{equation}
\langle r^2\rangle^\pi_V
  =\frac{6}{F_\pi^2}
      \left[2\Delta L_9 
           -\frac{1}{32\pi^2}\left[\frac{1}{6}+\frac{2}{9}Y(3)+\frac{M_\pi^2}{18M_K^2}(r+1)Y(3)\right]\right]
    +\langle r^2\rangle^\pi_V e^\pi_V\,,
\end{equation}
where we have introduced the  scale-independent combination $\Delta L_9=L_9^r(\mu)-\hat{L}_9^r(\mu)$, with:
\begin{equation}
\hat{L}_9^r(\mu)=\frac{1}{32\pi^2}
         \left[\frac{1}{6}\log\frac{\mo_\pi^2}{\mu^2}
              +\frac{1}{12}\log\frac{\mo_K^2}{\mu^2}\right] \,,
\end{equation}
and 
\begin{equation}
 e^\pi_V = \frac{6}{F_\pi^4}\frac{(\Re F_\pi^2F_\pi^V)_1}{\langle r^2\rangle^\pi_V}\,.
\end{equation}

The pion electromagnetic square radius is well-determined, and it is expected to suffer only mildly from higher-order corrections being an observable involving pions. We will thus use this observable to express $L_9$:
\begin{equation} \label{eq:deltaL9}
\Delta L_9=\frac{F_\pi^2}{12} \langle r^2\rangle^\pi_V [1-e^\pi_V]
        +\frac{1}{32\pi^2}\left[\frac{1}{12}+\frac{1}{9}Y(3)+\frac{M_\pi^2}{36M_K^2}(r+1)Y(3)\right]\,.
\end{equation}
Equivalent relations for other LECs, namely $L_{4,5,6,8}$ will be discussed later in eqs.~(\ref{eq:l4})-(\ref{eq:l8}).
$L_9(M_\rho)$ can thus be estimated as a function of $r,Y(3)$ and NNLO 
remainder. For instance, if we take $Y(3)=1$ and $r=2M_K^2/M_\pi^2-1$ (corresponding to LO estimates 
holding in the case of a fast convergence)
and the central experimental value~\cite{PDG}:
\begin{equation}
 \langle r^2\rangle^\pi_V = 0.451\pm 0.031\ {\rm fm}^2\,,
 \end{equation}
\noindent we obtain~\footnote{In this article, we take the following numerical values for the pseudoscalar masses, the pion decay constant and the renormalisation scale:
\begin{equation}
M_\pi=0.13957{\ \rm Gev}\,, \quad
M_K=0.4957{\ \rm Gev}\,, \quad
M_\eta=0.5478{\ \rm Gev}\,,\quad
      F_\pi=0.0922{\ \rm Gev}\,,\quad
       \mu=0.770{\ \rm Gev}\,.
\end{equation}}
$L_9^r(M_\rho)=6.77\cdot 10^{-3}$ in the ball park of usual estimates of 
this LEC, such as
$(6.9\pm 0.7)\cdot 10^{-3}$ at $O(p^4)$ in Ref.~\cite{chpt-su3}, 
 $(5.93\pm 0.43)\cdot 10^{-3}$ at $O(p^6)$ in Ref.~\cite{ffactors-em-p6}. On the other hand,
if $Y(3)$ tends to zero while the radius and its NNLO remainder remain finite, we see that $L_9$ becomes very large (we will see later that the other electromagnetic radii remain also finite in this limit).

\subsection{Masses and decay constants}

Exactly as in the previous section, we can write 
the identities for $F_P^2$ and $F_P^2 M_P^2$ in terms of
$r,X(3),Z(3)$, the $O(p^4)$ LECs $L_4,L_5,L_6,L_8$ and remainders:
\begin{eqnarray} \label{eq:fpi}
&&F_\pi^2 =
  F_\pi^2 Z(3)+M_\pi^2 Y(3)
           \left[8(r+2)L_4^r+8L_5^r\right]\\
&&\qquad      -\frac{1}{32\pi^2}M_\pi^2Y(3)
           \left[4\log\frac{{\mo}_\pi^2}{\mu^2}
            +(r+1)\log\frac{{\mo}_K^2}{\mu^2}\right]
 +F_\pi^2{e}_\pi\,,
\nonumber
\\ \label{eq:fka}
&&F_K^2 =
  F_\pi^2 Z(3)
+M_\pi^2 Y(3)
           \left[8(r+2)L_4^r
                     +4(r+1)L_5^r\right]\\
\nonumber
&&\qquad      -\frac{1}{32\pi^2}M_\pi^2Y(3)
      \Bigg[\frac{3}{2}\log\frac{{\mo}_\pi^2}{\mu^2}
            +\frac{3}{2}
         (r+1)\log\frac{{\mo}_K^2}{\mu^2}
            +\frac{1}{2}(2r+1)
                 \log\frac{{\mo}_\eta^2}{\mu^2}
           \Bigg]
 +F_K^2{e}_K\,,
\nonumber
\end{eqnarray}
\begin{eqnarray}
&&F_\pi^2M_\pi^2=
  F_\pi^2 M_\pi^2 X(3)
+M_\pi^4 [Y(3)]^2
           \left[16(r+2)L_6^r+16L_8^r\right]
\\
&&\qquad      -\frac{1}{32\pi^2}M_\pi^4[Y(3)]^2
           \Bigg[3\log\frac{{\mo}_\pi^2}{\mu^2}
            +(r+1)\log\frac{{\mo}_K^2}{\mu^2}
            +\frac{1}{9}(2r+1)
                   \log\frac{{\mo}_\eta^2}{\mu^2}
           \Bigg]\Bigg\}
 + F_\pi^2M_\pi^2{d}_\pi\,,
\nonumber
\end{eqnarray}
\begin{eqnarray}
 \label{eq:fkamka}
&&F_K^2M_K^2 =
  \frac{1}{2}(r+1)
    \Bigg\{F_\pi^2M_\pi^2 X(3)
+M_\pi^4 [Y(3)]^2
           \left[16(r+2)L_6^r
                     +8(r+1)L_8^r\right]
\\
&&\qquad
  -\frac{1}{32\pi^2}M_\pi^4[Y(3)]^2
           \Bigg[\frac{3}{2}\log\frac{{\mo}_\pi^2}{\mu^2}
            +\frac{3}{2}
                (r+1)\log\frac{{\mo}_K^2}{\mu^2}
            +\frac{5}{18}(2r+1)
                   \log\frac{{\mo}_\eta^2}{\mu^2}
           \Bigg]\Bigg\}
 +F_K^2 M_K^2 {d}_K\,.
\nonumber
\end{eqnarray}

The pion and kaon masses are well known experimentally. 
As far as the decay constants are concerned,
$F_\pi$ and $F_K$ are accessible at a high precision through leptonic decays 
($\pi_{l2}$ and $K_{l 2}$ respectively~\cite{PDG}) 
which provide in the framework of the Standard Model (SM)~\cite{Fnet}:
\begin{equation}
|V_{us}/V_{ud}| \times F_K/F_\pi=0.2758\pm 0.0005 \,,
\end{equation}
which can be combined with the very accurate determination of the first element of the CKM matrix
$V_{ud}$ from super-allowed $0^+ \to 0^+$ nuclear beta decays~\cite{ht09} 
\begin{equation}
V_{ud}=0.97425 \pm 0.00022\,,
\label{eq:vud}
\end{equation}
the unitarity of the CKM matrix and the smallness of the $|V_{ub}|$ matrix element to get:
\begin{equation} 
F_\pi\big|_{SM}=92.2 \pm 0.3 {\rm MeV} \, \qquad F_K/F_\pi\big|_{SM} =1.192 \pm 0.006\,,
\label{eq:decsm}
\end{equation}
In the following we take  the value of $F_\pi$ in eq.~(\ref{eq:decsm}), expecting 
the deviation from this SM determination to be rather small. On the other hand,
we will not fix the value of $F_K/F_\pi$, keeping it as a free parameter of 
the fit to the lattice data. A deviation from its value eq.(\ref{eq:decsm}) 
would hint at physics beyond the Standard Model contributing to flavour-changing charged currents, in addition to the usual $V-A$ term from the $W$ bosons.

Similarly to the case of $L_9$ and the electromagnetic square radius of the pion and 
following~\cite{resum,DescotesGenon:2007bs} we will then
invert the relationships eqs.~(\ref{eq:fpi})-(\ref{eq:fkamka}) in order to 
reexpress the four NLO LECs in terms of $X(3),Z(3),r$, the pion and kaon 
masses, the pion decay constant and the ratio $F_K/F_\pi$:
\begin{eqnarray} 
\label{eq:deltal4}
Y(3)\Delta L_4 &=& \frac{1}{8(r+2)}\frac{F_\pi^2}{M_\pi^2}
  [1-\eta(r)-Z(3)-e]\,,
\\ 
\label{eq:deltal5}
Y(3)\Delta L_5 &=& \frac{1}{8}\frac{F_\pi^2}{M_\pi^2}
  [\eta(r)+e']\,, 
\end{eqnarray}
\begin{eqnarray}
\label{eq:deltal6}
Y^2(3)\Delta L_6 &=& \frac{1}{16(r+2)}\frac{F_\pi^2}{M_\pi^2}
  [1-\epsilon(r)-X(3)-d]\,,
\\ 
\label{eq:deltal8}
Y^2(3)\Delta L_8 &=& \frac{1}{16}\frac{F_\pi^2}{M_\pi^2}
  [\epsilon(r)+d']\,.
\end{eqnarray}
$\Delta L_i=L_i^r(\mu)-\hat{L}_i(\mu)$ combine the 
(renormalized and quark-mass independent) 
constants $L_{4,5,6,8}$ and chiral logarithms so that they are
independent of the renormalisation scale $\mu$: 
\begin{eqnarray}
32\pi^2\hat{L}_4(\mu)
          &=& \frac{1}{8}
	  \log\frac{\mo_K^2}{\mu^2}
  -\frac{1}{8(r-1)(r+2)}
  \left[(4r+1)\log \frac{\mo_K^2}{\mo_\pi^2} 
      + (2r+1)\log \frac{\mo_\eta^2}{\mo_K^2}  \right] 	 
\label{eq:l4} \,,\\
32\pi^2 \hat{L}_5(\mu) &=& \frac{1}{8}
    \left[\log\frac{\mo_K^2}{\mu^2}+2\log\frac{\mo_\eta^2}{\mu^2}\right]
+\frac{1}{8(r-1)}
    \left[3\log\frac{\mo_\eta^2}{\mo_K^2}+5\log\frac{\mo_K^2}{\mo_\pi^2}\right]
\label{eq:l5} \,,\\
32\pi^2\hat{L}_6(\mu) &=& \frac{1}{16}\left[
       \log\frac{\mo_K^2}{\mu^2} 
       + \frac{2}{9}\log \frac{\mo_\eta^2}{\mu^2}
         \right] 
 -\frac{1}{16}\frac{r}{(r+2)(r-1)} \left[ 3 \log
          \frac{\mo_K^2}{\mo_\pi^2}  + \log \frac{\mo_\eta^2}{\mo_K^2} \right]
\label{eq:l6}\,, \\
32\pi^2 \hat{L}_8(\mu) &=& \frac{1}{16}
    \left[\log\frac{\mo_K^2}{\mu^2}+\frac{2}{3}\log\frac{\mo_\eta^2}{\mu^2}\right]
+\frac{1}{16(r-1)} \left[ 3 \log
          \frac{\mo_K^2}{\mo_\pi^2}  + \log \frac{\mo_\eta^2}{\mo_K^2} \right]
\label{eq:l8}\,.
\end{eqnarray}

The four equalities eqs.(\ref{eq:deltal4}-\ref{eq:deltal8}) are exact, since they are a mere reexpression of the
bare chiral series for $F_\pi^2$, $F_K^2$, $F_\pi^2M_\pi^2$ and $F_K^2 M_K^2$.
$d,d'$ and $e,e'$ are combinations of
 remainders associated with the chiral expansions of $\pi,K$ masses
and decay constants respectively:
\begin{eqnarray}
 d=\frac{r+1}{r-1}d_\pi - \left(\epsilon(r)+\frac{2}{r-1}\right)d_K \, , & &\qquad d'=d-d_\pi\,,\\
 e=\frac{r+1}{r-1}e_\pi - \left(\eta(r)+\frac{2}{r-1}\right)e_K \, ,& &\qquad e'=e-e_\pi\,.
\end{eqnarray}
$d',e'$ are quantities of $O(mm_s)$ whereas $d,e$ scale like $O(m_s^2)$.
In addition, the right hand-side of these equations involves the $r$-dependent functions:
\begin{equation} \label{funcr}
\epsilon(r) = 2\frac{r_2-r}{r^2-1}, \quad
r_2= 2\left(\frac{F_KM_K}{F_{\pi}M_{\pi}}\right)^2 -1\sim 36\,,\qquad
\eta(r)=\frac{2}{r-1}\left(\frac{F_K^2}{F_\pi^2}-1\right)\,.
\end{equation}

The properties of these equations, and in particular, the fact that they lead 
to a resummation of $L_4$ and $L_6$ contributions once inserted in the chiral 
expansion of other observables, were discussed at length in 
refs.~\cite{DescotesGenon:2007bs,resum,ordfluc}. Note that contrary to 
the case of $L_9$, these LECs always appear multiplied by powers of $Y(3)$ in chiral series,
since they correspond to operators with one or two powers of the scalar source in the chiral Lagrangian, and arise in chiral expansions always multiplied by one or two powers of $B_0$. Therefore, when $Y(3)$ tends to zero, $\hat{L}_i$ will exhibit a logarithmically divergent behaviour (due to the tadpole logarithms) which will not affect observables though.

\section{Kaon electromagnetic form factors} \label{sec:kaon}

The method described in the previous section can easily be generalized to other observables.
Of particular interest are the kaon electromagnetic form factors  and the $K \pi$ form factor which will be discussed in the following sections.
 
\subsection{Definition}

The kaon vector form factors~\cite{ffactors,ffactors-em-p6} are defined as:
\begin{equation}
\langle K^+|j_\mu|K^+\rangle = (p+p')^\mu F^{K^+}_V(t)\,, \qquad
\langle K^0|j_\mu|K^0\rangle = (p+p')^\mu F^{K^0}_V(t)\,,
\end{equation}
with the same convention as in the case of the pion electromagnetic form factor. All of
them are associated with the $P$-wave projection of the crossed channel. 
Following the discussion in \ref{sec:basics},
we expect $F_K^2 F_V^{K^+}$ and $F_K^2 F_V^{K^0}$ to have good convergence properties
away from the singularities (opening thresholds\ldots).
Expanding these form factors and reexpressing some couplings in terms of
$r,Y(3)$ and $Z(3)$, we obtain the bare expansion of the vector form factors: 
\begin{eqnarray}\label{eq:ka0vect-bare}
\frac{F_K^2}{F_\pi^2} F^{K^0}_V(t)
 &=&  -\frac{t}{192\pi^2F_\pi^2}\log\frac{\mo_K^2}{\mo_\pi^2}
      -\frac{1}{12 F_\pi^2}[t- 4 M_\pi^2 Y(3)]\bar{J}_{\pi\pi}(t)\\
\nonumber
&&\quad
        +\frac{1}{12 F_\pi^2}[t-2(r+1)M_\pi^2 Y(3)]\bar{J}_{KK}(t)
  +\frac{1}{F_\pi^2} \Re F_K^2F^{K^0}_V(t)\,,
\end{eqnarray}
\begin{eqnarray}
\label{eq:kaplusvect-bare}
\frac{F_K^2}{F_\pi^2}F^{K^+}_V(t)
 &=& Z(3) + 
       \frac{M_\pi^2}{F_\pi^2}[Y(3)]
          [8(r+2)L_4^r+4(r+1)L_5^r]\\
\nonumber
&&\quad
    -\frac{1}{32\pi^2}\frac{M_\pi^2}{F_\pi^2}Y(3)
          \Bigg[\frac{3}{2}\log\frac{\mo_\pi^2}{\mu^2}
                +\frac{3}{2}(r+1)\log\frac{\mo_K^2}{\mu^2}
                +\frac{1}{2}(2r+1)\log\frac{\mo_\eta^2}{\mu^2}
           \Bigg]\\
\nonumber
&&\quad
+\frac{t}{F_\pi^2}
   \Bigg[2L_9^r-\frac{1}{32\pi^2}
                 \left[\frac{1}{6}\log\frac{\mo_\pi^2}{\mu^2}
                      +\frac{1}{3}\log\frac{\mo_K^2}{\mu^2}
                      +\frac{1}{6}\right]\Bigg]\\
\nonumber
&&\quad
+\frac{1}{12 F_\pi^2}[t- 4 M_\pi^2 Y(3)]\bar{J}_{\pi\pi}(t)
        +\frac{1}{6 F_\pi^2}[t-2(r+1)M_\pi^2 Y(3)]\bar{J}_{KK}(t)\\
\nonumber
&&\quad
+\frac{1}{F_\pi^2}\Re F_K^2F^{K^+}_V(t)\,,
\end{eqnarray}
where $\Re  F_K^2F^{K^0}_V(t)$ and  $\Re  F_K^2F^{K^+}_V(t)$
 are polynomial functions of $t$ collecting
 remainders:
\begin{eqnarray}
\Re  F_K^2F^{K^0}_V(t)
  &=&  \frac{t}{F_K^2}(\Re  F_K^2F^{K^0}_V)_1  + O(t^2)\,,\\
\Re  F_K^2F^{K^+}_V(t)
  &=&  (\Re  F_K^2F^{K^+}_V)_0 
   + \frac{t}{F_K^2}(\Re F_K^2 F^{K^+}_V)_1  + O(t^2)\,,
\end{eqnarray}
with 
$(\Re F^K_V)_0 = O(m_q^2)$ and $(\Re F^K_V)_1 = O(m_q)$. We have divided the expressions of the form factors eqs.~(\ref{eq:kaplusvect-bare})-(\ref{eq:ka0vect-bare}) by a numerical factor $F_\pi^2$ for sole purpose of convenience, in order to deal with dimensionless quantities.

In the limit where all the observables are saturated by their leading order,  
the standard NLO chiral expansions of the vector form 
factors~\cite{ffactors,ffactors-em-p6} can be recovered by expanding the ratio 
$F_\pi^2/F_K^2$ at next-to-leading order and replacing the leading order masses by
the physical ones:
\begin{equation}
F^{K^0}_V(t)= -H_{\pi\pi}(t)+H_{KK}(t)+O(p^4)\,, \qquad
F^{K^+}_V(t)= F^\pi_V(t)+F^{K^0}_V(t)+O(p^4)\,,
\end{equation}
with $H_{PQ}$ defined as:
\begin{equation}
H_{PQ}(t)=\frac{1}{F_0^2}
   \left[
  \frac{1}{12}\left(t-2\Sigma_{PQ}+\frac{\Delta^2_{PQ}}{t}\right)
                \bar{J}_{PQ}(t)-\frac{\Delta_{PQ}^2}{3t}\bar{\bar{J}}_{PQ}(t)
                     -\frac{t}{6}k_{PQ}+\frac{t}{288\pi^2}\right]
          +\frac{2t}{3F_0^2}L_9^r\,,
\end{equation}
involving  $\Sigma_{PQ}=M_P^2+M_Q^2$ and $\Delta_{PQ}=M_P^2-M_Q^2$, and
$\bar{\bar{J}}_{PQ}(t)=\bar{J}_{PQ}(t)-t\bar{J}'_{PQ}(t)$.

\subsection{Kaon electromagnetic radii}

In a similar way to the pion form factor, the $K^+$ electromagnetic square radius is given by
\begin{equation}
\langle r^2\rangle^{K^+}_V
  =\frac{6}{F_K^2}
     \left[2\Delta L_9 -\frac{1}{32\pi^2}
          \left(\frac{1}{6}\log\frac{\mo_K^2}{\mo_\pi^2}+\frac{1}{6}
            +\frac{1}{9}Y(3)+\frac{M_\pi^2}{9M_K^2}(r+1)Y(3)\right)\right]
    + \langle r^2\rangle^{K^+}_V e^{K^+}_V\,,
\end{equation}
with the remainder:
\begin{equation}
e^{K^+}_V = \frac{6}{F_K^4}\frac{(\Re F_K^2 F_{K^+}^V)_1}{\langle r^2\rangle^{K^+}_V}\,.
\end{equation}
Replacing $\Delta L_9$ by its value in terms of the pion radius,
eq.(\ref{eq:deltaL9}), leads to the following  relation:
\begin{equation}
F_K^2 \langle r^2\rangle^{K^+}_V (1-e^{K^+}_V)
-F_\pi^2 \langle r^2\rangle^\pi_V (1-e^\pi_V)
 =\frac{1}{32\pi^2} 
    \left[-\log\frac{\mo_K^2}{\mo_\pi^2} +\frac{2}{3}Y(3)-\frac{M_\pi^2}{3M_K^2}(r+1)Y(3)\right]\,,
\label{eq:pikrad}
\end{equation}
where the right-hand side is a very small correction for any  reasonable value of $r$ and $Y(3)$, so that 
the electromagnetic square radius of the charged kaon is essentially predicted to be
$\langle r^2\rangle^{K^+}_V \simeq F_\pi^2/F_K^2 \times \langle r^2\rangle^{\pi}_V \simeq 0.32 \ {\rm fm}^2$. Two experiments have measured this radius, leading to the average~\cite{PDG}:
\begin{equation}
\langle r^2\rangle^{K^+}_V=0.314 \pm 0.035\ {\rm fm}^2\,.
\end{equation}
The square radius of the neutral kaon reads:
\begin{equation}
F_K^2\langle r^2\rangle^{K^0}_V (1-e^{K^0}_V)
 =\frac{1}{32\pi^2} 
    \left[-\log\frac{\mo_K^2}{\mo_\pi^2}+\frac{2}{3}Y(3)-
\frac{M_\pi^2}{3M_K^2}(r+1)Y(3)\right]
\label{eq:kneutrad}
\end{equation}
with the remainder:
\begin{equation}
e^{K^0}_V = \frac{6}{F_K^2}\frac{(\Re F_{K^0}^V)_1}{\langle r^2\rangle^{K^0}_V}\,,
\end{equation}
The current experimental average is~\cite{PDG}:
\begin{equation}
\langle r^2\rangle^{K^0}_V = -0.077 \pm 0.010 \ {\rm fm}^2\,.
\end{equation}
Eqs.~({\ref{eq:pikrad}}) and ({\ref{eq:kneutrad}}) yield the following relation between the
electromagnetic radii: 
\begin{equation}
\langle r^2\rangle^\pi_V (1-e^\pi_V)=\frac{F_K^2}{F_\pi^2} \left( \langle r^2\rangle^{K^+}_V (1-e^{K^+}_V)-\langle r^2\rangle^{K^0}_V (1-e^{K^0}_V)\right)
\,.
\end{equation}
which is fulfilled using the experimental values of the radii and the SM value of 
$F_K/F_\pi$ (the remainders must be on the large side
of their allowed value according to the dimensional estimation discussed in
Sec.~\ref{sec:remainders}). In principle, the knowledge of these remainders
and the determination of the radii with a high precision would allow us to 
determine $F_K/F_\pi$ accurately.

\section{$K\pi$ form factors} \label{sec:Kl3}

\subsection{Definition}

Among the quantities that can be determined from lattice simulations, one can single out the $K_{\ell 3}$ form factors defined as:
\begin{equation}
\sqrt{2}\langle K^+|\bar{u}\gamma_\mu s|\pi^0\rangle 
  = (p'+p)^\mu f_+(t)+(p'-p)^\mu f_-(t)\,.
\end{equation}
$f_+$ corresponds to $P$-wave projection of the $K_{\ell 3}$ transition, whereas its
$S$-wave comes from
\begin{equation}
f_0(t)=f_+(t)+\frac{t}{\Delta\kp}f_-(t)\,,
\label{eq:pikascal}
\end{equation}
where $\Delta_{PQ}=M_P^2-M_Q^2$.
Following the discussion in \ref{sec:basics},
$F_\pi F_K f_+$ and $F_\pi F_K f_-$ are expected to have good convergence 
properties away from the singularities (opening thresholds\ldots).
Exactly as before, their chiral expansions  can be expressed in terms of $r,X(3),Z(3)$, NLO low-energy constants 
($L_4$, $L_5$ and $L_9$) and  remainders.
Reexpressing $L_4$ and $L_5$ using eqs.~(\ref{eq:deltal4})-(\ref{eq:deltal5}) 
yields the following bare expansions of the $K_{\ell 3}$ form factors:
\begin{eqnarray}\label{eq:pikavect-bare1}
F_\pi F_K f_+(t)&=& \frac{F_\pi^2+F_K^2}{2} 
   + \frac{3}{2} [t M^r_{K\pi}(t) + t M^r_{K\eta}(t)
                  - L_{K\pi}(t) - L_{K\eta}(t)] \\
\nonumber
&& + 2 tL_9^r + F_\pi F_K d_+ + t e_+\,,
\end{eqnarray}
\begin{eqnarray} \label{eq:pikavect-bare2}
F_\pi F_K f_-(t) &=& \frac{F_K^2-F_\pi^2}{2}  
    -\frac{3}{2}(M^2_K-M^2_\pi) [M^r_{K\pi}(t) + M^r_{K\eta}(t)]
     \\
\nonumber
&&
  +\frac{1}{4}K_{K\pi}(t) \left[5(t-M_\pi^2-M_K^2)
                                  +\frac{3}{2}(r+3)M_\pi^2Y(3)\right]\\
\nonumber
&& -\frac{1}{4}K_{K\eta}(t)\left[3(t-M_\pi^2-M_K^2)
                                  +\frac{1}{2}(r+3)M_\pi^2Y(3)\right]\\
\nonumber
&& - 2(M^2_K-M^2_\pi) L_9^r  + F_\pi F_K (d_--d_+) + t (e_--e_+) \,,
\end{eqnarray}
where $d_\pm=O(m_q^2)$ and $e_\pm=O(m_q)$ combine the remainders from the form factors and the decay constants:
\begin{eqnarray}
F_\pi F_K d_+&=&   (\Re F_\pi F_K f_+)_0 - \frac{F_\pi^2 e_\pi +F_K^2 e_K}{2}\,, \\
F_\pi F_K (d_--d_+)&=&  (\Re F_\pi F_K f_-)_0 + \frac{F_\pi^2 e_\pi -F_K^2 e_K}{2}\,, \\
F_\pi F_K e_+&=&  (\Re F_\pi F_K f_+)_1\,,\\
F_\pi F_K (e_--e_+)&=&  (\Re F_\pi F_K f_+)_1\,,
\end{eqnarray}
where the remainder $F_\pi F_K (\Re f_\pm)(t)$ are defined as before.
When performing our fits to lattice data, we will also express $L_9$ in terms 
of the pion radius using  eq.~(\ref{eq:deltaL9}).
Inserting eqs.~(\ref{eq:pikavect-bare1})-(\ref{eq:pikavect-bare2}) into 
eq.~(\ref{eq:pikascal})
leads to the following expression for the scalar form factor:
\begin{eqnarray}\label{eq:pikascal-bare}
F_\pi F_K f_0(t)&=& \frac{F_K^2+F_\pi^2}{2}+\frac{t}{\Delta_{K\pi}} \frac{F_K^2-F_\pi^2}{2}
   - \frac{3}{2} L_{K\pi}(t) - \frac{3}{2} L_{K\eta}(t) 
   \\
&&
  +\frac{t}{4\Delta_{K\pi}}K_{K\pi}(t) \left[5(t-M_\pi^2-M_K^2)
                                  +\frac{3}{2}(r+3)M_\pi^2Y(3)\right]\nonumber\\
&&
  -\frac{t}{4\Delta_{K\pi}}K_{K\eta}(t)\left[3(t-M_\pi^2-M_K^2)
                                  +\frac{1}{2}(r+3)M_\pi^2Y(3)\right]\nonumber\\
&& + (F_\pi F_K d_+ + t e_+)\left(1-\frac{t}{\Delta_{K\pi}}\right)
 + (F_\pi F_K d_- + t e_-)\frac{t}{\Delta_{K\pi}}\,.\nonumber
\end{eqnarray}

In the limit where all expansions are saturated by their leading-order contribution, the well-known expression for the vector form factor is recovered:
\begin{equation}
f_+^{K\pi}(t)=1+\frac{3}{2}H_{K\pi}(t)+\frac{3}{2}H_{K\eta}(t)+O(p^4)\,,
\end{equation}
as well as that for $f_-^{K\pi}$~\cite{ffactors}.

\subsection{Callan-Treiman point, its soft kaon analog and the form factor
at zero momentum transfer}

According to the Callan-Treiman theorem~\cite{CT}, in the soft-pion limit ($p'^2=M_\pi^2=0$), the scalar form factor at $t=\Delta_{K\pi} \equiv M_K^2-M_\pi^2$ (Callan-Treiman point) should be equal to $F_K/F_\pi$.  
This implies that $F_K F_\pi f_0(\Delta_{K\pi})-F_K^2$ vanishes in the $N_f=2$ 
chiral limit $m_u=m_d=m\to 0$. There is a soft-kaon analog of this theorem holding at
$t=\tilde \Delta_{K\pi} \equiv -\Delta_{K\pi}$, stating that
$F_K F_\pi f_0(\tilde \Delta_{K\pi})-F_\pi^2$ vanishes in the $N_f=3$ chiral 
limit. At these particular points, eq.~(\ref{eq:pikascal-bare}) reads: 
\begin{eqnarray}\label{eq:CT1}
&&F_\pi F_K f_0(\Delta_{K\pi})= F_K^2
   - \frac{3}{2} L_{K\pi}(\Delta_{K\pi}) - \frac{3}{2} L_{K\eta}(\Delta_{K\pi}) 
   \\
&&\qquad
  +\frac{1}{4}K_{K\pi}(\Delta_{K\pi}) \left[-10M_\pi^2
                                  +\frac{3}{2}(r+3)M_\pi^2Y(3)\right]\nonumber
                                 \\
&&\qquad
  -\frac{1}{4}K_{K\eta}(\Delta_{K\pi})\left[-5M_\pi^2
                                  +\frac{1}{2}(r+3)M_\pi^2Y(3)\right]
+ F_\pi F_K d_- + \Delta_{K\pi} e_-\nonumber\,,\end{eqnarray}
\begin{eqnarray}
\label{eq:CT2}
&&F_\pi F_K f_0(-\Delta_{K\pi})= F_\pi^2
   - \frac{3}{2} L_{K\pi}(-\Delta_{K\pi}) - \frac{3}{2} L_{K\eta}(-\Delta_{K\pi}) 
   \\
&&\qquad
  -\frac{1}{4}K_{K\pi}(-\Delta_{K\pi}) \left[-10M_K^2
                                  +\frac{3}{2}(r+3)M_\pi^2Y(3)\right]\nonumber
                                  \\
&&\qquad
  +\frac{1}{4}K_{K\eta}(-\Delta_{K\pi})\left[-6M_K^2
                                  +\frac{1}{2}(r+3)M_\pi^2Y(3)\right]
+ F_\pi F_K (2d_+-d_-) - \Delta_{K\pi} (2e_+-e_-)\,.\nonumber
\end{eqnarray}

One can check explicitely that these expressions fulfill the Callan-Treiman theorem and its soft-kaon analog (the $K$ and $L$
contributions canceling each other) provided the following constraints on
the NNLO remainders 
\begin{equation}
d_-=O(m m_s)\,, \qquad e_-=O(m)\,,
\end{equation}
meaning that $d_-$ and $e_-$ are suppressed compared to $d_+$ and $e_+$.

We can define the discrepancies from the Callan-Treiman theorem(s):
\begin{equation}
\Delta_{CT}=f_0(\Delta_{K\pi})-\frac{F_K}{F_\pi}\,, \qquad
\tilde\Delta_{CT}=f_0(-\Delta_{K\pi})-\frac{F_\pi}{F_K}\,.
\end{equation}
These NLO quantities can be expressed from eqs.~(\ref{eq:CT1})-(\ref{eq:CT2}), embedding the 
fact that $\Delta_{CT}$ is $1/r$-suppressed compared to $\tilde\Delta_{CT}$.
For comparison, these quantities have been calculated in standard $\chi$PT at one-loop order
in the isospin limit~\cite{gl85}:
\begin{equation}
\Delta_{CT}= -3.5 \cdot 10^{-3}\,,\qquad
\tilde\Delta_{CT}=0.03\,.
\end{equation} 
It has in fact been 
shown in refs.~\cite{Bernard:2006gy,bops09} that a precise assessment of the scalar form factor at the Callan-Treiman points could probe physics beyond the Standard Model in the strange quark sector, in particular right-handed couplings of 
quarks to W bosons. The pioneering work~\cite{Bernard:2006gy} led to a reanalysis of 
$K_{\ell 3}$ data by several collaborations \cite{NA48mu,KLOEmu,KTeVmu},
which at present show a good/marginal agreement with the Standard Model except for the NA48
collaboration~\cite{NA48mu} exhibiting a $4.5\sigma$ deviation still unsolved.

The $K_{\ell 3}$ vector form factor at zero momentum transfer is another quantity of interest.
Indeed, the measurement of $K_{\ell 3}$ decays can be analysed in the framework of the Standard Model to determine the product 
$|V_{us} f_+(0)|$, and thus the CKM matrix element $|V_{us}|$. 
A recent fit to $|V_{ud}|$ (from super-allowed  $0^+ \to 0^+$ nuclear decays), $|V_{us}|f_+(0)$ (from $K_{\ell 3}$), and 
$|V_{us}/V_{ud}|F_K/F_\pi$ (from $\pi_{\ell 2}$ and $K_{\ell 2}$) together with the unitarity of the CKM matrix led to~\cite{Fnet}
\begin{equation} 
f_+(0)\big|_{SM}= 0.959 \pm 0.005 \,, 
\label{eq:fp0sm}
\end{equation}
and a value of $F_K/F_\pi\big|_{SM}$ in full agreement with eq.(\ref{eq:decsm}) with a strong correlation between these two quantities.

Deviation of $f_+(0)$ from this value would be an indication of new physics, so that
this quantity plays a particularly important role to test the Standard Model in the light quark sector. A direct determination of these quantities on the lattice as well as  
a well-controlled method to extrapolate lattice data down to the physical quark masses are naturally crucial to get a proper assessment of the uncertainties (from statistical, but also systematic origins).

\section{Observables for lattice simulations at different quark masses} \label{sec:lattice}

As explained in the previous section, we can use the relations eqs.~(\ref{eq:deltal4})-(\ref{eq:deltal5}) (decay constants),
 eqs.~(\ref{eq:deltal6})-(\ref{eq:deltal8}) (masses),
 eq.~(\ref{eq:deltaL9}) (pion electromagnetic square radius)\ldots to express NLO LECs in terms of $r$, $X(3)$, $Z(3)$, 
accurately measured observables and remainders. These relations can be inserted in the chiral expansions of 
other observables (such as kaon or $K_{\ell 3}$ form factors, or meson-meson 
scattering), which can be used to constrain $r$, $X(3)$ and $Z(3)$. For each new 
observable, one or several remainders are introduced, which are assumed to be 
small but nevertheless limit the accuracy of the chiral series.

As indicated in the introduction, one can also consider lattice simulations, where the same 
observables are considered at different values of the quark masses. The interest is twofold. First, 
the lattice simulations probe the $m_s$-sensitivity of observables,  hard to estimate from continuum 
measurements, but with deep connection with the pattern of $N_f=3$ chiral symmetry breaking. Second, 
lattice simulations  [RBC/UKQCD~\cite{Allton:2008pn,Boyle:2007qe,Boyle:2010bh}, PACS-CS~\cite{Aoki:2008sm}]
have encountered difficulties in their fits of NLO $N_f=3$ chiral expansions. Let us remark that
the inclusion of NNLO 
terms for $K_{\ell 3}$ form factors  and $F_K/F_\pi$ to fits of the RBC/UKQCD data seems to solve 
convergence issues  for these particular quantities assuming no
Zweig-rule violation in the scalar sector~\cite{Bernard:2009ds}: a good $\chi^2$ is obtained with a rather good convergence of the chiral series with NNLO terms of the expected size 
 (this work did not discuss $F_\pi$ itself, for which the problems of convergence
 seem the most acute~\cite{Allton:2008pn,Mawhinney:2009jy}).
 
Fitting these data will offer us the opportunity to extract relevant information 
on chiral symmetry breaking and  to  check the consistency of our picture
concerning the numerical competition between LO and NLO terms.
We consider simulations with 3 dynamical flavours $(\tilde{m},\tilde{m},\tilde{m}_s)$ and denote $\tilde{X}$ the values for the lattice quantities (and $X$ the corresponding value for physical quark masses). We introduce the ratios:
\begin{equation}
p=\frac{\tilde{m}_s}{m_s} \,,\qquad q=\frac{\tilde{m}}{\tilde{m}_s}\,,
\end{equation}
in addition to the ratio of physical quark masses $r$ and the chiral parameters arising in the leading-order Lagrangian in
eqs.~(\ref{eq:LOLECs}) and (\ref{eq:y3}).

\subsection{Masses and decay constants}

Proceeding as before in this new setting, we obtain the following
 expansions for the decay constants:
\begin{eqnarray} \label{eq:fpilatt}
&&\frac{\tilde{F}_\pi^2}{F_\pi^2} =
  Z(3)+\frac{M_\pi^2}{F_\pi^2}pqrY(3)
           \left[8\left(\frac{1}{q}+2\right)L_4^r+8L_5^r\right]\\
\nonumber
&&\qquad      -\frac{M_\pi^2}{F_\pi^2}\frac{1}{32\pi^2}pqrY(3)
           \left[4\log\frac{\tilde{\mo}_\pi^2}{\mu^2}
            +\left(\frac{1}{q}+1\right)\log\frac{\tilde{\mo}_K^2}{\mu^2}\right]
 +\frac{\tilde{F}_\pi^2}{F_\pi^2}\tilde{e}_\pi\,,
\nonumber
\end{eqnarray} 
\begin{eqnarray} 
 \label{eq:fkalatt}
&&\frac{\tilde{F}_K^2}{F_\pi^2} =
  Z(3)
+\frac{M_\pi^2}{F_\pi^2}pqrY(3)
           \left[8\left(\frac{1}{q}+2\right)L_4^r
                     +4\left(\frac{1}{q}+1\right)L_5^r\right]\\
\nonumber
&&\qquad      -\frac{M_\pi^2}{F_\pi^2}\frac{1}{32\pi^2}pqrY(3)
      \Bigg[\frac{3}{2}\log\frac{\tilde{\mo}_\pi^2}{\mu^2}
            +\frac{3}{2}
         \left(\frac{1}{q}+1\right)\log\frac{\tilde{\mo}_K^2}{\mu^2}
            +\frac{1}{2}\left(\frac{2}{q}+1\right)
                 \log\frac{\tilde{\mo}_\eta^2}{\mu^2}
           \Bigg]
 +\frac{\tilde{F}_K^2}{F_\pi^2}\tilde{e}_K\,,
\nonumber
\end{eqnarray}
where the LO contributions to the simulated pseudoscalar masses are involved:
\begin{equation}
\tilde\mo_\pi^2=pqr M_\pi^2 Y(3)\,, \qquad \tilde\mo_K^2=\frac{pqr}{2}\left(\frac{1}{q}+1\right) M_\pi^2Y(3)\,,
 \qquad \tilde\mo_\eta^2=\frac{pqr}{3}\left(\frac{2}{q}+1\right)M_\pi^2 Y(3)\,,
\label{eq:kring}
\end{equation}
and $\tilde{e}_P$ are remainders of $O(\tilde{m}_q^2)$ ($\tilde{m}_q$ 
denotes either $\tilde {m}_s$ or $\tilde{m}$). We have divided by the physical value
of $F_\pi^2$ in order to deal with dimensionless quantities.
In a similar way, we obtain the bare expansions of the masses:
\begin{eqnarray}
\label{eq:mpifit}
&&\frac{\tilde{F}_\pi^2\tilde{M}_\pi^2}{F_\pi^2M_\pi^2} =
  pqr\Bigg\{X(3)
+\frac{M_\pi^2}{F_\pi^2}pqr[Y(3)]^2
           \left[16\left(\frac{1}{q}+2\right)L_6^r+16L_8^r\right]
\\
&&\qquad      -\frac{M_\pi^2}{F_\pi^2}\frac{1}{32\pi^2}pqr[Y(3)]^2
           \Bigg[3\log\frac{\tilde{\mo}_\pi^2}{\mu^2}
            +\left(\frac{1}{q}+1\right)\log\frac{\tilde{\mo}_K^2}{\mu^2}
            +\frac{1}{9}\left(\frac{2}{q}+1\right)
                   \log\frac{\tilde{\mo}_\eta^2}{\mu^2}
           \Bigg]\Bigg\}
 + \frac{\tilde{F}_\pi^2\tilde{M}_\pi^2}{F_\pi^2M_\pi^2}\tilde{d}_\pi\,,
\nonumber\end{eqnarray}
\begin{eqnarray}\label{eq:fkamkalatt}
&&\frac{\tilde{F}_K^2\tilde{M}_K^2}{F_\pi^2M_\pi^2} =
  \frac{pqr}{2}\left(\frac{1}{q}+1\right)
    \Bigg\{X(3)
+\frac{M_\pi^2}{F_\pi^2}pqr[Y(3)]^2
           \left[16\left(\frac{1}{q}+2\right)L_6^r
                     +8\left(\frac{1}{q}+1\right)L_8^r\right]
\\
&&\qquad
  -\frac{M_\pi^2}{F_\pi^2}\frac{1}{32\pi^2}pqr[Y(3)]^2
           \Bigg[\frac{3}{2}\log\frac{\tilde{\mo}_\pi^2}{\mu^2}
            +\frac{3}{2}
                \left(\frac{1}{q}+1\right)\log\frac{\tilde{\mo}_K^2}{\mu^2}
            +\frac{5}{18}\left(\frac{2}{q}+1\right)
                   \log\frac{\tilde{\mo}_\eta^2}{\mu^2}
           \Bigg]\Bigg\}
 +\frac{\tilde{F}_K^2\tilde{M}_K^2}{F_\pi^2M_\pi^2}\tilde{d}_K\,,
\nonumber
\end{eqnarray}
where $\tilde{d}_P$ are remainders of $O(\tilde{m}_q^2)$.
 We have divided by the physical value
of $F_\pi^2 M_\pi^2$ in order to deal with dimensionless quantities.
As explained before, we use eqs.~(\ref{eq:deltal4})-(\ref{eq:deltal8}) to express the (mass-independent) chiral couplings $L_{4,5,6,8}$ in terms of $r,X(3),Z(3)$ and the physical masses and decay constants.


\subsection{$K_{\ell 3}$ form factors}

We obtain for the lattice vector form factor:
\begin{eqnarray}\label{eq:pikavect-lattice}
\tilde{F}_\pi \tilde{F}_K \tilde{f}_+(t)&=& \frac{\tilde{F}_\pi^2+\tilde{F}_K^2}{2} 
   + \frac{3}{2} [t \tilde{M}^r_{K\pi}(t) + t \tilde{M}^r_{K\eta}(t)
                  - \tilde{L}_{K\pi}(t) - \tilde{L}_{K\eta}(t)] \\
\nonumber
&& + 2 tL_9^r + \tilde{F}_\pi \tilde{F}_K \tilde{d}_+ + \tilde{t} e_+\,,
\end{eqnarray}
and the scalar form factor:
\begin{eqnarray}\label{eq:pikascal-lattice}
\tilde{F}_\pi \tilde{F}_K \tilde{f}_0(t)&=& \frac{\tilde{F}_K^2+\tilde{F}_\pi^2}{2}+\frac{t}{\tilde{\Delta}_{K\pi}} \frac{\tilde{F}_K^2-\tilde{F}_\pi^2}{2}\\
&&
   - \frac{3}{2} \tilde{L}_{K\pi}(t) - \frac{3}{2} \tilde{L}_{K\eta}(t) \nonumber\\
&&
  +\frac{t}{4\tilde{\Delta}_{K\pi}}\tilde{K}_{K\pi}(t) \left[5(t-\tilde{M}_\pi^2-\tilde{M}_K^2)
               +\frac{3}{2}\left(\frac{1}{q}+3\right)pqr M_\pi^2Y(3)\right]\nonumber\\
&&
  -\frac{t}{4\tilde\Delta_{K\pi}}\tilde{K}_{K\eta}(t)\left[3(t-\tilde{M}_\pi^2-\tilde{M}_K^2)
                                  +\frac{1}{2}\left(\frac{1}{q}+3\right)pqrM_\pi^2Y(3)\right]
\nonumber\\
&& + (\tilde{F}_\pi \tilde{F}_K \tilde{d}_+ + t \tilde{e}_+)
   \left(1-\frac{t}{\tilde{\Delta}_{K\pi}}\right)
 + (\tilde{F}_\pi \tilde{F}_K \tilde{d}_- + t \tilde{e}_-)\frac{t}{\tilde{\Delta}_{K\pi}}\,,\nonumber
\end{eqnarray}
where $\tilde{L}_{PQ}$, $\tilde{K}_{PQ}$, $\tilde{M}_{PQ}$ are evaluated 
with the leading-order pseudoscalar masses at the simulated quark masses using 
eq.~(\ref{eq:kring}), apart from the $\bar{J}_{PQ}$ function which is evaluated at the simulated ("physical") pion and kaon masses 
using eqs.~(\ref{eq:mpifit})-(\ref{eq:fkamkalatt}). In the above formulae, the decay constants on the 
right-hand side arise  from the reexpression of $L_4$ and $L_5$, and should be 
understood as a short-hand notation of the full expressions in 
eqs.~(\ref{eq:fpilatt})-(\ref{eq:fkalatt}).
For the vector form factor, we can trade $L_9$ for the pion electromagnetic square radius using eq.~(\ref{eq:deltaL9}).

\subsection{Remainders}\label{sec:remainders}

The expressions for the simulated masses, decay constants, form factors  and 
electromagnetic square  radius involve unknown remainders. These remainders
collect all the contributions coming from NNLO, NNNLO and higher orders. They
can be evaluated by resonance saturation~\cite{RS}, involving a hadronic scale 
$\Lambda_H$ only mildly affected by the actual value of the quark masses 
 (mass of the $\rho,K^*\ldots$).
In order to keep track of the scaling of the remainders with the quark masses, 
we take the following NNLO estimates which involves the hadronic scale at the fourth power:
\begin{eqnarray} \label{eq:rem1}
d,e,d_K,e_K,d_+=O\left(\frac{M_K^4}{\Lambda_H^4}\right)\,,
&\qquad& 
e_+=O\left(\frac{F_\pi^2 M_K^2}{\Lambda_H^4}\right)\,, \qquad
e_\pi^V=O\left(\frac{6}{\langle r^2\rangle^\pi_V}\frac{M_K^2}{\Lambda_H^4}\right)\,,\\
d',e',d_-=O\left(\frac{2M_\pi^2M_K^2}{\Lambda_H^4}\right)\,,
&\qquad& 
e_-=O\left(\frac{2F_\pi^2 M_\pi^2}{\Lambda_H^4}\right)\,,\\
d_\pi=d-d'\,, &\qquad & e_\pi=e-e'\,, \label{eq:rem2}
\end{eqnarray}
where $M_\pi^2$ and $M_K^2$ follow the known dependence of the remainders on $m$ 
and $m_s$, whereas $F_\pi^2$ is inserted when a dimensionful constant with 
no dependence on $m_q$ is required. 

We can use the known scaling of the remainders to perform their extrapolation to the simulated quark masses:
\begin{eqnarray}
\tilde{d}_\pi &=& p^2d-p^2qrd'\,,\qquad\qquad\qquad
\tilde{d}_K   = \left(\frac{F_K M_K}{F_\pi M_\pi}\right)^2 p^2\frac{r+1}{2}
              \left(d-\frac{r+1}{2}qrd'\right)\,,\\
\tilde{e}_\pi &=& p^2e-p^2qre'\,,\qquad\qquad\qquad
\tilde{e}_K   =  \left(\frac{F_{K}}{F_\pi}\right)^2 p^2\left(e-\frac{r+1}{2}qre'\right)\,,\\
\tilde{d}_+ &=& p^2d_+\,,\qquad
\tilde{e}_+ = p e_+\,,\qquad
\tilde{d}_- = p^2qrd_-\,,\qquad
\tilde{e}_- = pqr e_-\,.
\end{eqnarray}

\section{Fit to lattice values} \label{sec:fit}

\subsection{Data and parameters}

We are now in a position to build the $\chi^2$ function to be minimised for the sets of data that we will consider. The inputs are the following ones, as recalled in App.~\ref{app:inputs}:
\begin{itemize}
\item $\tilde{F}_\pi^2, \tilde{F}_K^2, \tilde{F}_\pi^2 \tilde{M}_\pi^2,  \tilde{F}_K^2 \tilde{M}_K^2$ for known values of the quark masses $(\tilde{m},\tilde{m}_s)$ (RBC/UKQCD and PACS-CS)
\item $\tilde{F}_\pi \tilde{F}_K \tilde{f}_+$ and $\tilde{F}_\pi \tilde{F}_K \tilde{f}_0$ for 
several transfer momenta (RBC/UKQCD)
\item We consider the  quantities  given by these collaborations
corresponding to light quark masses and small momenta ("Subset" fit) where chiral
perturbation theory is valid.
\end{itemize}

In Appendix B we will also consider all the quantities available even though
some points at higher masses and momenta might certainly be outside the region of validity of
Re$\chi$PT ("Total" fit)  to check the stability of our results as well as
to illustrate the interest of having more data points
to determine the remainders more accurately. Obviously, we would
need more points for lower pion and kaon masses.

The uncertainties on these quantities were obtained by combining the uncertainties in quadrature: no correlation between the various observables is provided in the articles of both collaborations, and we have only the statistical errors (no estimate of the systematic uncertainties was available for the quantities of interest here).
The parameters entering the fit are:
\begin{itemize}
\item Quantities from the leading-order chiral Lagrangian $X(3)$, $Z(3)$, $r$
\item NNLO remainders: $d, d', e, e'$ (in all cases), $d_+, e_+, d_-, e_-, e_\pi^V$ (for a fit to $K_{\ell 3}$ decay constants),
\item The value of the ratio of decay constants $F_K/F_\pi$,
\item The value of $p_{ref}=\tilde{m}_{s,ref}/m_s$ for a lattice set of reference, providing the equivalence between lattice and physical quark masses (when possible).
\end{itemize}
The last quantity is estimated by both collaborations, but we found it interesting to keep this parameter free in the fit, in order to take partially into account systematic effects related to lattice spacing. Since the quark masses are expressed in a mass-independent scheme involving only multiplicative renormalisation, we can determine the value of $p=\tilde{m}_s/m_s$ for any lattice set once we know $p$ for a given reference set using
$\tilde{m}_s=(\tilde{m}_s/\tilde{m}_{s,ref})\cdot p_{ref}$.
We had to fix the last two parameters in the case of RBC/UKQCD "Subset" fits, due to the limited number of unitarity (unquenched) points available for the
masses and decay constants (only 2 different pairs of quark masses).

When computing the values of the observables from NLO chiral expansions, we  need the values of the masses and decay constants for the simulated quark 
masses  (for instance in the unitarity functions $\bar{J}$). In such a case, we computed
systematically the values of the decay constants and masses from their chiral expansions~(\ref{eq:fpilatt})-(\ref{eq:fkamkalatt}), rather than plugging in their "measured" values on the lattice. This distinction may have some importance for the $K_{\ell 3}$ form factors eqs.~(\ref{eq:pikavect-bare1}) and (\ref{eq:pikascal-bare}), where we have reexpressed $L_4$ and $L_5$ in terms of $\tilde{F}_\pi^2$ and $\tilde{F}_K^2$, but where the latter quantities stand for their chiral expansion in terms of LECs (the fits to RBC/UKQCD would be slightly improved compared to the ones presented here if we used the measured values of $\tilde{F}_\pi^2$ and $\tilde{F}_K^2$ rather than the computed ones).

In addition, the mass $\tilde{M}_\eta$ and decay constant
$\tilde{F}_\eta$ of the $\eta$ are needed 
for the evaluation of the loop integral $\bar{J}_{PQ}$ 
(and related unitarity functions). They are obtained at a sufficient accuracy for such purposes
using the two following LO formulae reminiscent of the Gell-Mann-Okubo relation:
\begin{equation}
\tilde{F}_\eta^2=\frac{4}{3}\tilde{F}_K^2-\frac{1}{3}\tilde{F}_\pi^2\,,\qquad
\tilde{F}_\eta^2\tilde{M}_\eta^2=\frac{4}{3}\tilde{F}_K^2\tilde{M}_K^2-\frac{1}{3}\tilde{F}_\pi^2\tilde{M}_\pi^2\,.
\end{equation}

We constrain the remainders in the ranges indicated in
Table~\ref{table:remainders}. Once the (MINUIT-powered) fit has converged, we can 
estimate a large body of quantities: NLO LECs, $N_f=2$ chiral order parameters, 
values of the $K_{\ell 3}$ scalar form factor at zero momentum transfer,
at the Callan-Treiman point and its soft kaon analog, test of the convergence of the series. We have propagated the errors exploiting the 
covariance matrix provided by MINOS, assuming that all uncertainties follow a 
Gaussian distribution.

\subsection{Discussion} \label{sec:discussion}

Our results are summarised in Table~\ref{table:results}. The first series of rows corresponds to the outcome of the fit,  whereas the lower rows are quantities derived from the results of the fit (LO LECs, NLO LECs, quantities in the $N_f=2$ chiral limit, $K_{\ell 3}$ quantities, relative fraction of LO/NLO/remainders contributions at the minimum for several observables), and the last row is the $\chi^2$ per degree of freedom.

The "Subset" fit of RBC/UKQCD results includes only a limited number of data points for the masses and decay constants, which forces us to fix one of 
the parameters of the fit, namely the simulated strange quark mass~\footnote{Letting all parameters free gives 
comparable results for the central values, but some parameters get very large uncertainties,
larger than their allowed range. Propagating the errors in such a situation would be meaningless, and reporting the results of this fit would not provide much more information than the constrained fit that we present here.} and to impose some bounds on the size of the higher-order remainders, based on a simple estimate from resonance saturation described in App.~\ref{app:remainders}. Indeed, some of these remainders are pushed to the limits of their range 
when the set of data is too small, because there is not enough 
information for MINUIT to choose a particular value for these remainders (keeping them free would lead to a larger contribution from higher orders and to a further decrease of the leading-order contribution).
The situation improves when higher masses and momenta are included (see 
the fits to "Total" sets presented in App.~\ref{sec:total}), so that the remainders remain small, within the
window discussed in App.~\ref{app:remainders}. 
It is interesting to notice that the results are consistent with 
the fits to PACS-CS data ("Subset" or "Total"): $r$ is close to 25 (i.e  close to the ratio $2M_K^2/M_\pi^2-1$ even though its value was left free in our 
framework), the quark condensate remains around $X(3)\sim 0.5$, the squared decay constant is $Z(3)\sim 0.6$, leading to a value for the LO term of the squared pion mass 
$Y(3)\sim 0.8$.

The ratio of decay constants $F_K/F_\pi$ (left free in 
our fit) comes out slightly larger (smaller) for PACS-CS (RBC/UKQCD) than its Standard Model value eq.(\ref{eq:decsm}) in the
fits of the masses and decay constants. We obtain also values 
of simulated strange quark masses and of the physical strange quark mass in 
good agreement with the results obtained by the two collaborations 
(the discrepancy between RBC/UKQCD and PACS-CS is due to the different choice 
of renormalisation procedure, which explains the low value obtained by the 
PACS-CS collaboration~\cite{Aoki:2009ix}).

\begin{table}[h!]
\begin{equation*}
\begin{array}{|c|c|c|}
\hline
\phantom{xx}&&\\[-2.4ex]
 & {\rm PACS-CS\ Subset}& {\rm RBC/UKQCD\ Subset}\\[0.15ex]
 &  {\rm Without\ } K_{\ell3}& {\rm With\ } K_{\ell3} \\
\hline
\phantom{xx}&&\\[-2.4ex]
r          & 26.5\pm 2.3 & 23.2\pm 1.5 \\
X(3)      & 0.59\pm 0.21 & 0.20 \pm 0.14 \\
Y(3)     & 0.90\pm 0.22  & 0.43 \pm 0.30\\
Z(3)     & 0.66\pm 0.09 & 0.46 \pm 0.04 \\
F_K/F_\pi   &  1.237 \pm 0.025 & 1.148 \pm 0.015\\
{\rm Rem.\ at\ limit}  &  {\rm none} &  d,e \\
\tilde{m}_{s,ref}/m_s 
          & 1.24\pm 0.08 & 1.15^\star\\[0.15ex]
\hline
\phantom{xx}&&\\[-2.4ex]
m_s(2\ {\rm GeV}) [{\rm MeV}] &  70 \pm 4 & 107\\
m(2\ {\rm GeV})   [{\rm MeV}]  &  2.6\pm 0.3 & 4.6\pm 0.3  \\
B_0(2\ {\rm GeV}) [{\rm GeV}] &  3.34\pm 1.18 & 0.92\pm 0.67\\
F_0 [{\rm MeV}]               &  74.8\pm 4.9 & 62.2\pm 2.5\\[0.15ex]
\hline
\phantom{xx}&&\\[-2.4ex]
 L_4(\mu)\cdot 10^3   &  -0.1\pm 0.2 & 2.4\pm 2.0  \\
 L_5(\mu)\cdot 10^3   &  1.8\pm 0.4 & 1.8\pm 1.6\\
 L_6(\mu)\cdot 10^3   & 0.1\pm 0.4 & 4.7\pm 7.1\\
 L_8(\mu)\cdot 10^3   &  0.8\pm 0.7 & 4.4\pm 7.1 \\
 L_9(\mu)\cdot 10^3  & \times & 4.4\pm 2.8 \\[0.15ex]
 \hline
\phantom{xx}&&\\[-2.4ex]
X(2)  & 0.90\pm 0.01 & 0.90\pm 0.02  \\
Y(2)  & 1.04\pm 0.02 & 1.00\pm 0.03  \\
Z(2)  & 0.87\pm 0.02 & 0.90\pm 0.02\\
B (2\ {\rm GeV}) [{\rm GeV}] & 3.83\pm 0.50  & 2.09\pm 0.19 \\
F [{\rm MeV}]                &  85.8 \pm 0.7  & 87.7\pm 0.8 \\
\bar\ell_3       & 5.0\pm 2.1  & -0.6\pm 3.7 \\
\bar\ell_4       & 4.5\pm 0.5  & 3.3\pm 0.5\\
\Sigma/\Sigma_0  &  1.51\pm 0.51& 4.52\pm 2.83\\
B/B_0           &  1.15\pm 0.26 & 2.28\pm 1.39 \\
F/F_0           &  1.15\pm 0.08  & 1.41\pm 0.06\\[0.8ex]
\hline
\phantom{xx}&&\\[-2.4ex]
f_+(0)          &  1.004\pm 0.149 & 0.985\pm 0.008   \\
\Delta_{CT}\cdot 10^3 & \times & -0.2\pm 12.1 \\
\Delta'_{CT}\cdot 10^3& \times & -126\pm 104 \\[0.15ex]
\hline
\phantom{xx}&&\\[-2.4ex]
\langle r^2\rangle_V^{K^+} [{\rm fm}^2] & \times &
0.248\pm 0.156  \\
\langle r^2\rangle_V^{K^0} [{\rm fm}^2] & \times &
-0.027 \pm 0.106 \\[0.15ex]
\hline
\phantom{xx}&&\\[-2.4ex]
F_\pi^2  
        & 0.66 + 0.22 + 0.12 &  0.45 + 0.69 - 0.14 
\\
F_K^2   
        & 0.44 + 0.48 + 0.08  & 0.34 + 0.76 - 0.10 
 \\
F_\pi^2M_\pi^2 
        & 0.60 + 0.30 + 0.10 & 0.20 + 0.95 - 0.15
\\
F_K^2M_K^2
        & 0.42 + 0.50 + 0.08 & 0.14 + 0.97 - 0.11\\
F_\pi F_K f_+(0)
       & \times &  0.40+ 0.75 - 0.15
\\[0.8ex]
\hline\phantom{xx}&&\\[-2.4ex]
\chi^2/N   &  0.9/3 & 4.4/8 \\[0.15ex]
\hline
\end{array}
\end{equation*}

\caption{Results of fits performed on the data from the PACS-CS~\cite{Aoki:2008sm} and
RBC/UKQCD~\cite{Allton:2008pn,Boyle:2007qe,Boyle:2010bh}
collaborations on pseudoscalar masses and decay constants, and $K_{\ell 3}$ form factors in the case of RBC/UKQCD. In all cases, we considered only data with light pions (Subset)
and only statistical errors are shown. In the RBC/UKQCD case, we fixed the lattice strange quark mass (marked with a star). 
The LECs are given at the scale $\mu =m_\rho$. In the PACS-CS case, the  
$K_{\ell 3}$ form factor at zero momentum transfer is a prediction of the fit
(with an error combining those obtained from the fit and the maximal contribution
allowed for the remainder from dimensional estimation). The penultimate set of rows collects the relative fractions of LO/NLO/remainders for
decay constants, masses and  $K_{\ell 3}$ form factor at vanishing 
transfer momentum (for RBC/UKQCD) at the minimum.}\label{table:results}
\end{table}

\clearpage

The decay constant in the $N_f=3$ chiral limit is found to be rather low, 
in agreement with other recent works~\cite{Bijnens:2009hy,Bernard:2009ds}. 
The pattern of $N_f=3$ chiral symmetry breaking (with low quark condensate 
and decay constant) is reflected by the values obtained for the low-energy 
constants $L_4$ and $L_6$, which are both positive and do not show any sign of 
Zweig suppression. Such large values of $L_4$ and/or $L_6$ have been obtained in several 
earlier works: dispersive analysis of scalar form factors~\cite{bachirl6}, dispersive treatment of $\pi K$ scattering~\cite{roypika},
 $J/\psi$ decay into a vector meson and two pseudoscalars~\cite{lamei06} (with a value of $L_6$ compatible with zero),
preliminary NNLO $N_f=3$ fits to pseudoscalar masses, decay constants, $K_{\ell 4}$ decay and $\pi K$ scattering data
\cite{Bijnens:2009hy}\ldots
Large values of $L_4$ and $L_6$ are known to induce a significant dependence 
of the chiral order parameter on the strange quark mass, and it is not 
surprising to witness a strong suppression from the $N_f=2$ chiral limit to the 
$N_f=3$ one. Let us notice that we obtain values for $N_f=2$ chiral order 
parameters~\footnote{The expressions for the $N_f=2$ order parameters $X(2)$, $Y(2)$, $Z(2)$ can be obtained by setting $\tilde{m}=0, \tilde{m}_s=m_s$ in the expressions for $\tilde{F}_\pi^2$ and $\tilde{F}_\pi^2 \tilde{M}_\pi^2$,
eqs.~(\ref{eq:fpilatt}) and (\ref{eq:mpifit}).
The matching expressions for $\bar\ell_{3,4}$ in terms of $L_{4,5,6,8}$ differ
from the usual ones~\cite{chpt-su3,Gasser:2009hr} by factors of $Y(2)$ and $Y(3)$ that are easy to recover.}
 which are in agreement with the hypothesis of standard $\chi$PT 
confirming that this framework is indeed appropriate in the $N_f=2$ sector
as shown by the recent data on $K_{\ell 4}$
decays~\cite{Batley:2007zz,:2009nv}, though this does not seem to be the case for $N_f=3$.
 The other LECs $L_{5,8,9}$ have values in agreement with 
conventional estimates (this was expected in particular for $L_9$ since
our framework induces modifications that are only sub-leading for vector 
quantities  such as the pion electromagnetic form factor).

Let us notice some specificities of our treatment of the chiral expansions.
Tadpoles diagrams generate chiral logarithms of the form
$M_P^2\log(M_P^2/\mu^2)$ which can prove quite troublesome to fit.
For instance, the RBC/UKQCD collaboration~\cite{Allton:2008pn,Boyle:2007qe,Boyle:2010bh} finds a better agreement of their data on decay constants with polynomial fits than with chiral series. In our treatment, these chiral logarithms always involve the leading-order 
mass $\mo_P^2$. Therefore, the limit of a small $Y(3)$ tames the chiral logarithms in our expansions, so that these logarithms become 
hard to distinguish from a polynomial at the numerical level on the range of masses where $\chi$PT could be valid. 
Furthermore large contributions from NLO LECs, and in particular from $L_4$ and $L_6$ as just discussed, will enhance the quadratic dependence on the quark masses  and thus our chiral expressions will mimic a polynomial dependence on the quark masses that cannot be reproduced in the more usual treatment of chiral expansions.
These mechanisms could explain why chiral logarithms are often difficult to identify in lattice data, in addition to other effects 
(heavy strange quark mass, lattice systematics\ldots).

The $K_{\ell 3}$ form factor at zero momentum transfer, $f_+(0)=f_0(0)$, 
involves only LECs related to decay constants and masses, eq.~(\ref{eq:pikavect-bare1}). In principle, it can be predicted from a fit of the latter quantities up to
the determination of the remainder $d_+$. We quote the corresponding
results in Table~\ref{table:results}, where the central value for $f_+(0)$  corresponds to
remainders set to zero. The uncertainty on this quantity includes the
maximal size allowed for the remainders $d_+$ based on dimensional
estimate (see Table~\ref{table:remainders}), as well as the uncertainties
coming from the parameters of the fit. Clearly,  
the NNLO remainder $d_+$  hinders any accurate 
determination of $f_+(0)$, 
unless their value is also precisely determined from the fit, which is possible
once data on $K_{\ell 3}$ form factors themselves is included 
(Table~\ref{table:results}). 
The values obtained for $f_0(0)$ are somewhat larger than the Standard Model value 
eq.(\ref{eq:fp0sm}), as well as those
obtained from the RBC/UKQCD collaboration using different forms for 
the extrapolation in quark masses~\cite{Boyle:2007qe,Boyle:2010bh}. This 
illustrates the importance of the mass extrapolation for lattice simulations 
at the level of accuracy aimed at currently. A 
particular attention was paid in ref.~\cite{Boyle:2010bh}
 to the structure of the chiral expansion of $f_+(0)=1+f_2+f_4$, where 
$f_2$ is the NLO contribution, which involves only a combination of chiral 
logarithms \emph{divided by} $F_0^2$:
\begin{equation}\label{eq:f2}
 f_2=-\frac{3}{256\pi^2 F_0^2}\left[(M_K^2+M_\pi^2)\,h\!\!\left(\frac{M_\pi^2}{M_K^2}\right)+(M_K^2+M_\eta^2)\,h\!\!\left(\frac{M_\eta^2}{M_K^2}\right)\right]\,,
 \qquad h(x)=1+\frac{2x}{1-x^2}\log x\,.
 \end{equation}
$f_2$ is often said to be free from LECs and thus known precisely from Chiral 
Perturbation Theory. This statement is not totally correct for the 
following reasons.
One usually assumes that the value of $F_0$ is 
close to that of
$F_\pi$, so that it can be replaced in actual calculations by the physical value of the pion decay constant leading to the estimate $f_2\simeq -0.023$. Since
the difference between the two quantities is a higher-order effect, 
one can always perform this replacement. However, one has to determine how large the
NNLO term $f_4$ can be with such a prescription, and consequently how well the chiral
series for $f_+$ converges. If $F_0$ is significantly lower than $F_\pi$ as indicated not 
only by our fit, but by other recent estimates~\cite{Bijnens:2009hy,Bernard:2009ds},  
the convergence is expected to be rather slow, forcing us to treat
the NLO contribution to $f_+(0)$ more carefully. We advocated that correlators of vector and 
axial currents yields observables with good convergence properties, selecting
 $F_\pi F_K f_0(0)$.  In this case, we should 
replace $F_0^2$ by $F_\pi F_K$ in the evaluation of eq.~(\ref{eq:f2}), as can 
be checked in our expression for $f_+$, eq.~(\ref{eq:pikavect-bare1}).

Once the $K_{\ell 3}$ form factors are included in our fits, $L_9$
can be determined even though the fit does not constrain  for this 
particular LEC tightly. In Table~\ref{table:results}, the deviations from the 
Callan-Treiman relation at $t=\Delta_{K\pi}$ and its soft kaon analog at 
$-\Delta_{K\pi}$ are given. Their values are of the expected size for
$SU(N_f)$ chiral-symmetry breaking quantities for $N_f=2,3$ flavours respectively, 
and thus compatible with the one obtained in standard $\chi$PT.  The  values of the square radii 
of the charged and neutral kaons, also shown have rather large uncertainties and are thus
within the experimental error bars.

In the last lines of our tables, we have indicated for each fit the contribution from LO, NLO and remainders to pseudoscalar decay constants and masses for values of the parameters at the minimum of the fit. We can see that the series converge well on overall (remainder much smaller than LO+NLO), but that the LO term is far from saturating the series. The values of $Y(3)$ obtained is smaller than 1, reducing the contribution from chiral logarithms compared to that from the NLO LECs.
We can compare these results with those from a fit of the same observables, where the NLO and higher contributions (chiral logarithms $\mu_P$, LECs $L_i$, remainders) are computed replacing $2mB_0$, $(m+m_s)B_0$ and $F_0$ by the physical pion and kaon masses and the pion decay constant.
This is exactly equivalent to performing the same fit as before with the following replacements in the NLO and higher-order contributions:
\begin{eqnarray} \label{eq:LOvalues}
&& r \to 2\frac{M_K^2}{M_\pi^2}-1, \qquad
q \to \frac{\tilde{M}_\pi^2}{2\tilde{M}_K^2-\tilde{M}_\pi^2},\qquad
p \to \frac{2\tilde{M}_K^2-\tilde{M}_\pi^2}{2M_K^2-M_\pi^2},\qquad
Y(3)\to 1,\\
&& \eta(r)\to \eta(r_0), \qquad \epsilon(r)\to \epsilon(r_0)-2X(3)\frac{r-r_0}{r_0^2-1},\qquad
 \log\frac{\mo^2_P}{\mu^2}\to \log\frac{M^2_P}{\mu^2},
\end{eqnarray}
both for the observables that we consider, eqs.~(\ref{eq:fpilatt})-(\ref{eq:pikascal-lattice}), and the equations allowing the determination of $L_{4,5,6,8,9}$, eqs.~(\ref{eq:deltaL9}) and (\ref{eq:l4})-(\ref{eq:l8}).
For PACS-CS, this leads to $\chi^2/N=1.1/3$ (compared to our result 0.9/3), with very similar values for the fundamental parameters $r,X(3),Y(3),Z(3)$. For RBC/UKQCD, the fitting procedure yields $\chi^2/N=9.5/8$ (compared to our result 4.4/8), with much more uncertain values of the fundamental parameters (e.g., $r=14.9\pm 12.1$, $X(3)=0.30\pm 0.26$, $Y(3)=0.68\pm 0.60$). This is not particularly surprising since our fits to the PACS-CS data led to values of $r$ and $Y(3)$ in good agreement with eq.~(\ref{eq:LOvalues}), but not the RBC/UKQCD ones. The corresponding convergence of the pseudoscalar masses and decay constants is then (the relative contribution for LO, NLO and higher orders is given here):
\begin{equation*}
\begin{array}{ccrclcrcl}
\ {\mathrm{PACS-CS\ [NLO\ phys.\ masses]}} && 
  F_\pi^2&:& 0.64+0.26+0.10 ,&  & F_\pi^2 M_\pi^2&:& 0.67+0.24+0.09,\\
&&  F_K^2&:& 0.42 + 0.51 +0.07&  & F_K^2M_K^2&:& 0.50+0.44+0.06,\\
\ {\mathrm{RBC/UKQCD\ [NLO\ phys.\ masses]}} && 
  F_\pi^2&:& 0.45 + 0.70 - 0.15 &  & F_\pi^2 M_\pi^2&:& 0.31+0.81-0.12 , \\
&&  F_K^2&:& 0.34 + 0.76 - 0.10 ,&& F_K^2 M_K^2&:& 0.15+0.94-0.11.
\end{array}
\end{equation*}
There is no saturation of the series by their leading order. We see that our formulae yields results that are in good agreement with those obtained after reexpressing the NLO contributions in terms of $F_\pi, M_\pi, M_K$ in the PACS-CS case, where $Y(3)$ is close to 1. On the other hand, when $Y(3)$ is not close to 1 (for instance in the RBC/UKQCD case), our formulae provide more efficient and accurate fits (lower $\chi^2$, smaller error bars). From a more methodological point of view, we avoid a perturbative reexpression of LECs in terms of $F_\pi, M_\pi, M_K$ in a regime where it is not justified.

These trends can be compared interestingly with the fits done by the lattice collaborations themselves, with a different treatment of the chiral series than ours. For instance, the MILC collaboration~\cite{Bazavov:2009tw} observed from fits with staggered chiral perturbation theory that $M_\pi^2$ received NNLO corrections of the same size as NLO contributions, canceling each other to a large extent, with small NNNLO corrections (the latter being taken as analytic in quark masses and lattice spacings), whereas $F_\pi$ exhibited no problems of convergence. On the other hand, the RBC/UKQCD collaboration~\cite{Allton:2008pn} experienced difficulties in fitting $F_\pi$ both in $N_f=2$ and $N_f=3$ theories. They also noticed that fits to $M_K^2$ and $F_K$ using the $N_f=3$ chiral expansion led to very significant NLO contributions (of order 50\%) when data up to the kaon mass scale was included, and they conclude that higher-order corrections could be very significant (up to 30\%).

At this point, we should emphasize that our framework does not contain any bias
concerning the size of $X(3)$, $Y(3)$ and $Z(3)$ or on the relative size of the
LO and NLO contributions. It is compatible with the usual assumptions that chiral series
of decay constants, squared masses\ldots are saturated by
their LO contribution, but it can also accommodate situations where there is a numerical competition between LO and NLO terms.
It turns out that the lattice data set from the RBC/UKQCD and PACS-CS collaborations favour values for the three quantities $X(3),Y(3), Z(3)$ smaller than 1, 
with a  $\chi^2$/d.o.f. which ranges from fairly good to 
excellent.
Our results confirm the difficulties 
reported by the two collaborations to fit $N_f=3$ NLO chiral expressions, and 
highlights the improvement provided by our Re$\chi$PT formulae for the extrapolations in quark masses of these quantities.
As a further check, we have performed fits where we have taken that the physical masses (and not the LO ones) in the unitarity functions 
$J,K,L,M$ and the argument of the chiral logarithms (similarly to what was done in ref.~\cite{DescotesGenon:2007bs}). 
The quality and parameters of the fits are almost unchanged, and the outcome for the derived quantities is also very similar, meaning 
that the relevant issue is the proper choice of the "good observables" whose chiral series converge well.

\section{Conclusion}

Recent lattice simulations with light 2+1 dynamical fermions have encountered 
difficulties to fit their results for pseudoscalar masses, decay constants and form factors 
with chiral expansions obtained from $N_f=3$ Chiral Perturbation Theory at next-to-leading order. Such fits of poor quality can be related to the fact that chiral series are not saturated by their leading order, so that there is a numerical competition between leading-order contributions -- from the decay constant and/or the condensate in the $N_f=3$ chiral limit ($m_u=m_d=m_s=0$) -- and next-to-leading-order contributions -- in particular from $L_4$ and $L_6$, related to the Zweig-rule violation in the scalar sector, enhanced by $m_s$ and not accurately known. 

If there is such a competition, one must decide which observables are expected to have 
a good overall convergence (small higher-order contributions). According to the 
assumed equivalence of the $\chi$PT and QCD generating functionals at low energies, it seems reasonable to consider observables derived from correlators of axial and vector currents as well as 
pseudoscalar and scalar densities, as done here.
For these observables, one must treat chiral series with a particular care, avoiding the 
perturbative reexpression of LECs in terms of observables while neglecting higher orders (this 
can be easily done by introducing remainders corresponding to NNLO and higher contributions) and choosing how unitarity contributions should be treated to define the structure of the chiral expansion and its splitting into leading, next-to-leading and higher-order terms. Such a set of prescriptions was introduced some time ago under the name of Resummed Chiral Perturbation Theory.
In the present paper, we have recalled the basic ingredients of this framework
and applied it to observables related to pseudoscalar masses, decay constants, 
and kaon and pion form factors (electromagnetic and $K_{\ell 3}$ ones). This 
allowed us to illustrate how $O(p^4)$ LECs $L_{4,5,6,8,9}$ can be reexpressed 
in terms of the leading-order quantities $X(3), Z(3), r$ as well as 
experimental values of observables (pion and kaon decay constants and masses, 
square electromagnetic radius of the pion) and associated remainders.

Then we have turned to 2+1 lattice simulations where these observables were obtained for 
several sets of quark masses: PACS-CS (decay constants and masses only)~\cite{Aoki:2008sm} 
and RBC/UKQCD (decay constants, masses, $K_{\ell 3}$ form factors)~\cite{Allton:2008pn,Boyle:2007qe,Boyle:2010bh}. We performed fits to data corresponding only 
to light quark masses and small momenta, but checked the stability of our procedure by  considering also fits to all data available (unitary points). Since only statistical 
uncertainties (without correlations) are publicly available for each of the points, we performed naive fits with Gaussian errors, in order to determine the leading-order parameters of the chiral Lagrangian as well as  NNLO remainders and the ratio of decay constants.

The fits are generally of a good quality, with a good consistency when one compares subsets coming from the same collaboration. This allows one to determine the values of the LO quantities as well as the NNLO remainders, with a good accuracy in the case of PACS-CS, with a more limited precision for RBC/UKQCD because of the restricted number of low-mass points. One observes that:
\begin{itemize}
\item The decay constant and the quark condensate in the $N_f=3$ limit ($m_u=m_d=m_s=0$) are both small and suppressed compared to the $N_f=2$ case ($m_u=m_d=0$ and $m_s$ physical).
\item The low-energy constants $L_4$ and $L_6$ do not follow the Zweig rule suppression generally advocated to set them to zero.
\item The other low-energy constants $L_5$, $L_8$ and $L_9$ have values in good agreement with previous estimates.
\item The ratio of quark masses $r$ remains quite close to the most simple estimate from pseudoscalar masses.
\item $N_f=2$ chiral order parameters are in good agreement with the values extracted from $K_{\ell 4}$ decays.
\item When the sets of data are large enough, the NNLO remainders remain in the expected range from a naive dimensional estimate.
\item The expected numerical competition between LO and NLO chiral expansions indeed occurs for $F_\pi^2$, $F_K^2$, $F_\pi^2M_\pi^2$ and $F_K^2M_K^2$.
\end{itemize}
Beyond this description of the pattern of $N_f=3$ chiral symmetry breaking and its implication for the convergence of chiral expansions, we can also make a few predictions. The values obtained for the kaon electromagnetic radii are in good agreement with experimental data.
In the case of RBC/UKQCD, the value obtained for $f_+(0)$ with our fits is slightly larger than the ones quoted by the collaboration, relying on alternative formulae for the chiral expansion of the $K_{\ell 3}$ form factors. This has naturally an impact on the determination of $|V_{us}|$, considering the level of accuracy achieved in $K_{\ell 3}$ decays~\cite{Fnet}.

Lattice simulations are able to investigate the dependence of observables on quark masses, which makes them very valuable tools to investigate the chiral structure of QCD vacuum. Conversely, any improvement in our understanding of chiral symmetry breaking will help reducing systematics associated with chiral extrapolations in lattice determinations. In light of the discussion presented in this paper, it would be very helpful that more lattice collaborations present the dependence of their results on the quark masses and/or study alternative ways of performing the extrapolation down to physical quark masses in order to assess the related systematics precisely. It would also be very interesting to have lattice results for observables related to the scalar channel, and thus difficult to determine from experiment. For instance, the pion and kaon scalar form factors, simulated on the lattice and analysed in the framework of Resummed Chiral Perturbation, would provide an interesting complement to the present discussion.

\section*{Acknowledgments}

It is a pleasure to thank Damir Becirevic, Beno\^{\i}t Blossier, Andreas 
J\"uttner, Laurent Lellouch and Chris Sachrajda  for helpful discussions and 
comments on lattice 
simulations and Ulf-G. Mei{\ss}ner for careful reading of the manuscript. We would like to thank  
Enno Scholz particularly, for 
providing error estimates for combinations of RBC/UKQCD lattice data, allowing us to account for some 
correlations between the quantities discussed here.   
Work supported in part by EU Contract No.  MRTN-CT-2006-035482, \lq\lq FLAVIAnet''.

\appendix

\section{Lattice inputs} \label{app:inputs}

\subsection{RBC/UKQCD Collaboration}

We first consider the RBC/UKQCD Collaboration simulations with 2+1 dynamical flavours~\cite{Allton:2008pn,Boyle:2007qe,Boyle:2010bh} performed with domain-wall fermions at one lattice spacing $a^{-1}=1.729(28)$ GeV. The calculations are performed on two volumes,
$16^3\times 32$ and $24^3\times 64$ ($(2.74)^3 {\rm fm}^3$) (with a fifth dimension of length 16), at each quark mass, except the lightest mass which is only simulated
on the larger volume.  They performed a non-perturbative renormalisation to relate the lattice quark masses to those in the RI-MOM scheme.

The only points that we will use are those where the sea and valence  quark masses are identical. There are four sets corresponding to such a situation for pseudoscalar masses and decay constants in  Ref.~\cite{Allton:2008pn}, corresponding to $a(\tilde{m}^{lat}-m_{res})$ and $a(\tilde{m}^{lat}_s-m_{res})$ being respectively
$(0.005,0.040)$, $(0.010,0.040)$, $(0.020,0.020)$, $(0.030,0.030)$.
where $am_{res}=0.00315(2)$. The quark masses are given in the RI-MOM scheme, but they can be related to the $\bar{MS}$ scheme through a multiplicative factor
$\bar{m}(2{\rm\ GeV})=Z_m a^{-1}(a\tilde{m}^{lat})$ that drops in all the input quantities (which involve only ratio of quark masses in the same setting).

We obtain the following values expressed in units of $10^{-3}\ {\rm GeV}^{-2}$ for the pseudoscalar masses and decay constants~\cite{Allton:2008pn,Schultz:private}. The uncertainties here are purely statistical and do not  include those induced by the uncertainty on the value of the lattice spacing.
\begin{equation*}
\begin{array}{c|c|c|c|c|c|c}
{\rm Masses} & (p,q) &  F_\pi^2  & F_\pi^2 M_\pi^2 &  F_K^2 &    F_K^2 M_K^2 & {\rm Subset}\\
\hline
(0.005,0.040) & (1.15, 0.189)  & 10.98 \pm 0.16 &1.196 \pm 0.022 &   14.11 \pm 0.19 &  4.644 \pm 0.076 &\star\\
(0.010,0.040 )& (1.15,0.304) &    12.85\pm  0.16  & 2.249 \pm 0.036 & 15.59 \pm 0.18 & 5.730 \pm 0.082  &\star\\
(0.020,0.020) & (0.616,1)    &           15.58 \pm  0.34 & 4.851 \pm 0.107 & 15.58 \pm 0.34 & 4.851 \pm 0.107 &\\
(0.030,0.030) & (0.883,1) &                  17.82 \pm 0.36 & 8.038 \pm  0.166 & 17.82 \pm 0.36 & 8.038 \pm 0.166 &\\
\end{array}
\end{equation*}

In two papers~\cite{Boyle:2007qe,Boyle:2010bh}, the RBC/UKQCD collaboration investigated the $K_{\ell 3}$ form factors $f_0$ and $f_+$ using in particular twisted boundary conditions to obtain a sample transfer momenta, with the same two sets of values corresponding to nondegenerate masses $(a(\tilde{m}^{lat}-m_{res}),a(\tilde{m}^{lat}_s-m_{res}))=(0.005,0.040),(0.010,0.040)$.

The set with the lighter $u,d$ quark masses yields the following values: 
 \begin{equation*}
 \begin{array}{c|c|c|c|c|c}
 t &   60.7 &   59.87 &  38.1& 21.6 &  0.30 \\
 \hline
  F_\pi F_K f_0(t) &
12.68 \pm 0.17 & 12.73 \pm 0.17 & 12.49 \pm  0.17 & 12.32 \pm  0.17 &
12.15 \pm  0.16 \\
 F_\pi F_K f_+(t) &
\times & \times & 12.71 \pm 0.176 & 12.42\pm 0.175 & 12.15 \pm 0.17\\
{\rm Subset} & \star & \star & \star & \star & \star 
\end{array}
\end{equation*}
 \begin{equation*}
 \begin{array}{c|c|c|c|c}
 t &   -44.00 &  -129.3 &  -204.9 &  -389.2 \\
 \hline
  F_\pi F_K f_0(t) &
 11.68 \pm 0.21 & 10.95 \pm 0.32 &10.77 \pm 0.23 &  9.667 \pm 0.28\\
 F_\pi F_K f_+(t) & \times & \times & \times & \times\\
{\rm Subset} & \star & \star & & 
\end{array}
\end{equation*}

The set with the heavier value leads to the following values for the scalar form factor:
 \begin{equation*}
 \begin{array}{c|c|c|c|c|c}
 t & 35.42 & -90.51 & -195.3 & -205.0 & -385.2\\
 \hline
F_\pi F_K f_0(t) &  14.28 \pm 0.17 & 13.05 \pm  0.21 &
  11.64 \pm 0.38 & 12.89 \pm 0.40 & 11.83 \pm 0.75 \\
{\rm Subset} & \star & \star & \star & &
 \end{array}
 \end{equation*}

We have considered also a subset of data, indicated with stars, where the convergence of chiral series is expected to be particularly good. This amount to considering only non-degenerate $u,d,s$ quark masses, and to drop the points for $t\leq -0.2\ {\rm Gev}^{-2}$ [i.e., the two points for $f_0$ in ref.~\cite{Boyle:2007qe} corresponding to the lowest values of transfer momentum].

The values of the physical quark masses ($m$ and $m_s$) and the lattice spacing
are obtained by studying the dependence of the mass of $\pi$, $K$ and $\Omega$ hadrons on these three parameters and tuning
them to reproduce the physical hadron masses. If we call $\tilde{m}_{s,ref}$ the value of the strange quark mass corresponding to
the set (0.005,0.040), the RBC/UKQCD collaboration obtained $\tilde{m}_{s,ref}/m_s=1.150$. Considering the uncertainty associated with such a determination  (in particular the role played by the form of the chiral extrapolation used for $\pi$ and $K$), we will not assume this value in our fit, but rather include this quantity as a parameter of our fit, and scale the other ratios involving a simulated strange quark mass over the physical value.

Fits to the $N_f=2$ and $N_f=3$ NLO chiral series for pseudoscalar masses and decay constants were performed in ref.~\cite{Allton:2008pn}. It turned out that the $N_f=3$ chiral expansions led to rather poor fits (large $\chi^2$ per d.o.f), in particular for decay constants, unless they put stringent cuts on the values of quark masses where such expansions should hold. This led the authors in ref.~\cite{Allton:2008pn}
to perform fits to $N_f=2$ NLO chiral expansions.
In addition, in ref.~\cite{Mawhinney:2009jy}, NNLO $SU(2)$ chiral expansions were shown to 
have only a limited utility to extrapolate the data: many more data points would be needed to fix the size of the combinations of $O(p^6)$ counterterms involved
The results obtained in ref.~\cite{Allton:2008pn} that are relevant for our discussion are summarised in table~\ref{table:RBC}.

\begin{table}
\begin{equation*}
\begin{array}{|c|c|}
\hline
r & 28.8\pm 0.4\pm 1.6\\
\tilde{m}_{s,ref}/m_s & 1.150\\
F_K/F_\pi & 1.205\pm 0.018\pm 0.062\\
m_s(2\ {\rm GeV}) [{\rm MeV}]& 107.3\pm 4.4\pm 9.7\pm 4.9\\
m(2\ {\rm GeV})   [{\rm MeV}]& 3.72\pm 0.16\pm  0.33\pm 0.18\\
B (2\ {\rm GeV}) [{\rm GeV}]& 2.52\pm 0.11\pm 0.23\pm 0.12\\
F [{\rm MeV}]               & 81.2\pm 2.9\pm 5.7\\
\bar\ell_3 & 3.13\pm 0.33 \pm 0.24\\
\bar\ell_4 & 4.43\pm 0.14 \pm 0.77\\
\hline
\end{array}
\end{equation*}
\caption{Results obtained by the RBC/UKQCD collaboration in ref.~\cite{Allton:2008pn}.\label{table:RBC}}
\end{table}

In addition, two different values for $f_+(0)$ were obtained in refs.~\cite{Boyle:2007qe,Boyle:2010bh}
 from the same
gauge configurations, using either data for the scalar form factor or data
for both form factors, and applying 
a pole ansatz based on either $N_f=3$ or $N_f=2$ chiral
perturbation theory for $K_{\ell 3}$ form factors~\cite{Flynn:2008tg}:
\begin{equation}
f_+(0)=0.964\pm 0.033\pm 0.0034\pm 0.0014~\cite{Boyle:2007qe}\,,\qquad\qquad
f_+(0)=0.960(^{+5}_{-6})~\cite{Boyle:2010bh}\,.
\end{equation}

\subsection{PACS-CS collaboration}

The PACS-CS collaboration~\cite{Aoki:2008sm} has investigated the pseudoscalar 
masses and decay constants with a large sample of light quark masses, for one particular value of lattice spacing $a^{-1}=2.176(31)$ GeV, on a $32^3 \times 64$ 
volume. They used a non-perturbatively $O(a)$-improved Wilson quark action  and performed the renormalisation of quark masses perturbatively at one loop (with tadpole improvement), with the following results:
\begin{equation*}
\begin{array}{c|c|c|c|c|c|c}
(am_{ud}^{\bar{MS}},am_s^{\bar{MS}}) & (p,q) &  F_\pi^2  & F_\pi^2 M_\pi^2 &  F_K^2 &    F_K^2 M_K^2 & {\rm Subset}\\
\hline
(0.001,0.040) & (1.410,0.040) &
    10.19\pm 1.09 &  0.247 \pm 0.035 & 14.29 \pm 0.48 & 4.385 \pm 0.151 & \star \\
(0.006,0.041) & (1.456,0.138) &  
    11.51\pm 0.26 & 1.007\pm 0.031   &  15.49 \pm  0.22 & 5.459 \pm 0.088 & \star\\
(0.010,0.036) & (1.256,0.271) &
    12.48 \pm 0.21 & 1.846\pm 0.041 & 15.37 \pm 0.16 & 5.200 \pm 0.067 & \star\\
(0.011,0.042) & (1.519,0.260) &
     13.25 \pm 0.18 & 2.242\pm 0.036 & 16.83 \pm 0.23 & 6.791 \pm 0.096 &\\
(0.021,0.045) & (1.577,0.466) &
     17.23 \pm 0.73 & 5.595 \pm 0.239 & 19.87 \pm 0.65 & 10.11 \pm 0.33 &\\
(0.031,0.047) & (1.663,0.652) &
     19.09 \pm 0.51 & 9.397 \pm 0.254 & 21.01 \pm 0.58 & 13.08 \pm 0.37 &
 \end{array}
 \end{equation*}
Once again, the uncertainties are of purely statistical origin, and they do not include the uncertainty coming from the determination of the lattice spacing. We have also considered a subset of data, indicated with stars, where the convergence of the chiral series is expected to be particularly good. This amounts to considering the three lightest values of the pion masses.

\begin{table}[h!]
\begin{equation*}
\begin{array}{|c|c|}
\hline
r       & 28.8\pm 0.4\\
Y(3)   & 0.88\pm 0.01\\
Z(3)   & 0.76\pm 0.04\\
F_K/F_\pi & 1.189 \pm 0.020\\
\tilde{m}_{s,ref}/m_s & 1.19\\
m_s(2\ {\rm GeV}) [{\rm MeV}]& 72.72\pm 0.78\\
m(2\ {\rm GeV})   [{\rm MeV}]& 2.527\pm 0.047\\
B_0(2\ {\rm GeV}) [{\rm GeV}]& 3.869\pm 0.092 \\
F_0 [{\rm MeV}]               &  83.8\pm 6.4\\
 L_4(\mu)\cdot 10^3 & -0.06\pm 0.10\\
 L_5(\mu)\cdot 10^3 & 1.45\pm 0.07\\
 L_6(\mu)\cdot 10^3 & 0.03 \pm 0.05\\
 L_8(\mu)\cdot 10^3 & 0.61 \pm 0.04\\
Y(2) &  0.96\pm 0.01\\
Z(2) & 0.88\pm 0.01\\  
B (2\ {\rm GeV}) [{\rm GeV}]& 0.96\pm 0.01\\
F [{\rm MeV}]               & 88.2\pm 3.4\\
\bar\ell_3 & 3.14\pm 0.23\\
\bar\ell_4 & 4.04\pm 0.19\\
\Sigma/\Sigma_0 & 1.205\pm 0.014 \\
B/B_0           & 1.073\pm 0.055\\
F/F_0           & 1.065\pm 0.058\\
\hline
\end{array}
\end{equation*}
\caption{Results obtained by the CP-PACS collaboration with one-loop perturbative renormalisation and extrapolation to the physical limit~\cite{Aoki:2008sm}. The values for the quantities in the $N_f=2$ chiral limit correspond to $N_f=2$ fits to the so-called Range I with finite-size effects included.\label{table:PACS1}}
\end{table}

The values of the physical quark masses ($m$ and $m_s$) and the lattice spacing
are obtained by studying the dependence of the mass of $\pi$, $K$ and $\Omega$ hadrons on these three parameters and tuning
them to reproduce the physical hadron masses. If we call $\tilde{m}_{s,ref}$ the value of the strange quark mass corresponding to
the set (0.0016,0.0399), we obtain $\tilde{m}_{s,ref}/m_s=1.19$. Considering the uncertainty associated with such a determination (in particular the role played by the form of the chiral extrapolation used for $\pi$ and $K$), we will not assume this value in our fit, but rather include this quantity as a parameter of our fit, and scale the other ratios involving a simulated strange quark mass over the physical value. 

Fits to the $N_f=2$ and $N_f=3$ NLO chiral series for pseudoscalar masses and decay constants were performed in ref.~\cite{Aoki:2008sm}. It turned out that the $N_f=3$ chiral expansions led to rather poor fits, related to very significant NLO contributions compared to LO terms, in particular for the decay constants, related to large contributions from kaon loops. In other words, the dependence of these quantities on the strange quark mass seen in these simulations  is not accounted for properly by NLO $SU(3)$ chiral perturbation theory. This led the authors in ref.~\cite{Aoki:2008sm}
to perform fits to $N_f=2$ chiral expansions.
The results obtained in ref.~\cite{Aoki:2008sm} that are relevant for our discussion are summarised in table~\ref{table:PACS1}.

A latter article of the same collaboration~\cite{Aoki:2009ix} considered  simulations directly performed at the physical point
including non-perturbative renormalisation. This has induced a significant modification for the quark mass renormalisation factor,
becoming $Z_m=1.441(15)$ (non-perturbative) instead of $Z_m=1.114$ (one-loop perturbation theory) leading to an increase (decrease) in the values of quark masses (condensates) by a factor 1.30. This should be taken into account when
comparing the results obtained from the PACS and RBC/UKQCD sets in this article. The results obtained in ref.~\cite{Aoki:2009ix} that are relevant for our discussion are summarised in table~\ref{table:PACS2}. Since the simulation was performed at the physical point, there is no further information on LECs describing the pattern of $N_f=2$ and $N_f=3$ chiral symmetry breakings.

\begin{table}[h!]
\begin{equation*}
\begin{array}{|c|c|}
\hline
r       & 31.2\pm 2.7\\
F_K/F_\pi & 1.333 \pm 0.072\\
m_s(2\ {\rm GeV}) [{\rm MeV}]& 92.75\pm 0.58\pm 0.95\\
m(2\ {\rm GeV})   [{\rm MeV}]& 2.97\pm 0.28 \pm 0.03\\
\hline
\end{array}
\end{equation*}
\caption{Results obtained by the CP-PACS collaboration~\cite{Aoki:2009ix} with non-perturbative renormalisation and simulation at the physical point.\label{table:PACS2}}
\end{table}

\section{Dimensional estimate of the remainders}\label{app:remainders}

As indicated in eqs.~(\ref{eq:rem1})-(\ref{eq:rem2}), we can estimate the higher-order remainders assuming that they are dominated by NNLO contributions and using
resonance saturation. A typical order of magnitude for 
$O(m_s^2)$ remainders so that the chiral series converge is 10\% so that 
$\Lambda_H\simeq$ 0.8 GeV.  The corresponding 
size $\sigma$ of the remainders is given in table~\ref{table:remainders}, and the remainders
will be required to stay in the range $[-\sigma,\sigma]$ in our fits to lattice data if necessary. In the specific case of the electromagnetic square radius of the pion, we have combined the uncertainty on the experimental measurement of the square radius with the theory uncertainty on the remainder in quadrature (the range for the kaon radii would be the same).

\begin{table}[h]
\begin{center}
\begin{tabular}{|c|c|}
\hline
Remainder & $\sigma$\\
\hline
$d,e,d_+$ & 0.148 \\
$d',e',d_-$ & 0.024 \\
$e_+$ & 0.005 \\
$e_-$ &  0.001 \\
$e_\pi^V$ & 0.318\\
\hline
\end{tabular}
\caption{Size of the NNLO remainders allowed in our fits, based on a dimensional estimate. \label{table:remainders}}
\end{center}
\end{table}

\section{Fits with "Total" data sets} \label{sec:total}

In Sec.~\ref{sec:discussion}, we considered fits to both PACS-CS and RBC/UKQCD data restricted to the low-mass and low-momentum region ("Subset" data). We have also performed fits to the whole sets of data available ("Total" data), in order to test the stability of our results, and to illustrate the interest of having larger data sets to determine NNLO remainders in an accurate way.
We are aware that some of the data points considered here may stand outside the region of validity for Re$\chi$PT, but we found nevertheless interesting to provide these results, showing a good consistency with those obtained with "Subset" data.

\begin{table}[h!]
\begin{equation*}
\begin{array}{|c|c|c|}
\hline
\phantom{xx}&&\\[-1.4ex]
 {\rm Without\ } K_{\ell3} & {\rm RBC/UKQCD\ Total} &  {\rm PACS-CS\ Total}\\[0.8ex]
\hline
\phantom{xx}&&\\[-1.4ex]
r          & 25.8\pm 0.9    &   25.7\pm 0.9\\
X(3)      & 0.44\pm 0.03 & 0.48\pm 0.04\\
Y(3)     & 0.77\pm 0.06 & 0.76\pm 0.07\\
Z(3)     & 0.56\pm 0.04 & 0.63 \pm 0.05\\
F_K/F_\pi   & 1.214\pm 0.012 &  1.239 \pm 0.011\\
{\rm Rem.\ at\ limit}  & {\rm none}& {\rm none}\\
\tilde{m}_{s,ref}/m_s 
          & 1.12\pm 0.03 &  1.21 \pm 0.02\\[0.8ex]
\hline
\phantom{xx}&&\\[-1.4ex]
m_s(2\ {\rm GeV}) [{\rm MeV}] & 110\pm 3 & 72\pm 2\\
m(2\ {\rm GeV})   [{\rm MeV}]  & 4.3 \pm 0.1 &  2.8\pm 0.1 \\
B_0(2\ {\rm GeV}) [{\rm GeV}] & 1.75\pm 0.14 &  2.65\pm 0.20 \\
F_0 [{\rm MeV}]               & 69.2\pm 2.2 &  73.0\pm 2.5\\[0.8ex]
\hline
\phantom{xx}&&\\[-1.4ex]
 L_4(\mu)\cdot 10^3   & 0.7\pm 0.1  &  0.1\pm 0.2\\
 L_5(\mu)\cdot 10^3   & 1.8\pm 0.2  & 2.2\pm 0.2\\
 L_6(\mu)\cdot 10^3   & 0.8\pm 0.2 &  0.5\pm 0.3\\
 L_8(\mu)\cdot 10^3   & 1.2\pm 0.3  &  1.4\pm 0.3\\[0.8ex]
 \hline
\phantom{xx}&&\\[-1.4ex]
X(2)  & 0.90\pm 0.01  & 0.90\pm 0.01\\
Y(2)  & 1.02\pm 0.01  & 1.03\pm 0.01\\
Z(2)  & 0.88\pm 0.01  & 0.87\pm 0.01\\
B (2\ {\rm GeV}) [{\rm GeV}] & 2.34\pm 0.07 & 3.58\pm 0.11 \\
F [{\rm MeV}]                & 86.7\pm 0.3&  86.3\pm 0.3 \\
\bar\ell_3       & 3.1\pm 0.6  & 3.8\pm 0.6\\
\bar\ell_4       & 3.9\pm 0.2  & 4.2\pm 0.2\\
\Sigma/\Sigma_0  & 2.07\pm 0.15 & 1.88\pm 0.14\\
B/B_0           & 1.32\pm 0.10  & 1.35 \pm 0.11 \\
F/F_0           & 1.25\pm 0.04  & 1.18 \pm 0.04\\[0.8ex]
\hline
\phantom{xx}&&\\[-1.4ex]
f_+(0)          & 1.006\pm 0.149 &  1.011\pm 0.149\\
[0.8ex]
\hline
\phantom{xx}&&\\[-1.4ex]
F_\pi^2  
        & 0.56 + 0.54 - 0.10
        & 0.63 + 0.28 + 0.09
\\
F_K^2   
        & 0.38 + 0.69 - 0.07
        & 0.41 + 0.52 + 0.07
 \\
F_\pi^2M_\pi^2 
        & 0.44 + 0.67 - 0.11     
        & 0.48 + 0.50 + 0.02
\\
F_K^2M_K^2
        & 0.31 + 0.77 - 0.08
        & 0.33 + 0.65 + 0.02
\\[0.8ex]
\hline\phantom{xx}&&\\[-1.4ex]
\chi^2/N   & 13.6/7 & 13.8/15 \\[0.8ex]
\hline
\end{array}
\end{equation*}

\caption{Results of fits performed on the data from the
RBC/UKQCD~\cite{Allton:2008pn,Boyle:2007qe,Boyle:2010bh} and PACS-CS~\cite{Aoki:2008sm} 
collaborations on pseudoscalar masses and decay constants, considering all the available unquenched data (Total). Only statistical errors are shown. 
The LECs are given at the scale $\mu =m_\rho$. The  
$K_{\ell 3}$ form factor at zero momentum transfer is a prediction of the fit
(with an error combining those obtained from the fit and the maximal contribution
allowed for the remainder from dimensional estimation). 
The penultimate set of rows collects the relative fractions of LO/NLO/remainders for
decay constants and  masses at the minimum.}\label{table:resultstotal}
\end{table}

\begin{table}[h!]
\begin{equation*}
\begin{array}{|c|c|c|}
\hline
\phantom{xx}&&\\[-2.4ex]
{\rm With\ }K_{\ell3} &  {\rm RBC/UKQCD\ Total\ I} & {\rm RBC/UKQCD\ Total\ II}\\[0.15ex]
\hline
\phantom{xx}&&\\[-2.4ex]
r           & 24.9 \pm 0.6  & 25.2 \pm 0.9 \\
X(3)   & 0.43 \pm 0.03 & 0.42 \pm 0.03\\
Y(3)   & 0.80 \pm 0.05 & 0.78 \pm 0.06\\
Z(3)    & 0.53 \pm 0.03 & 0.54 \pm 0.04\\
F_K/F_\pi & 1.199 \pm 0.009 & 1.203\pm 0.011\\
{\rm Rem.\ at\ limit} & {\rm none} & {\rm none}\\
\tilde{m}_{s,ref}/m_s 
         & 1.15^\star & 1.14\pm 0.03\\[0.15ex]
\hline
\phantom{xx}&&\\[-2.4ex]
m_s(2\ {\rm GeV}) [{\rm MeV}] & 107    & 109\pm 3\\
m(2\ {\rm GeV})   [{\rm MeV}] & 4.3 \pm 0.1  & 4.3 \pm 0.1\\
B_0(2\ {\rm GeV}) [{\rm GeV}] & 1.80\pm 0.12 & 1.77\pm 0.14\\
F_0 [{\rm MeV}]                & 67.1\pm 1.9  & 67.6\pm 2.1\\[0.15ex]
\hline
\phantom{xx}&&\\[-2.4ex]
 L_4(\mu)\cdot 10^3   & 0.76\pm 0.10 & 0.75\pm 0.10\\
 L_5(\mu)\cdot 10^3   & 1.64\pm 0.12 & 1.71\pm 0.19\\
 L_6(\mu)\cdot 10^3   & 0.71\pm 0.13 & 0.76\pm 0.17\\
 L_8(\mu)\cdot 10^3  & 1.18\pm 0.22 & 1.19\pm 0.23\\
 L_9(\mu)\cdot 10^3  & 5.05\pm 2.25 & 5.08\pm 2.25\\[0.15ex]
 \hline
\phantom{xx}&&\\[-2.4ex]
X(2)  & 0.90\pm 0.01 & 0.90\pm 0.01\\
Y(2)  & 1.02\pm 0.01 & 1.02\pm 0.01\\
Z(2) & 0.88\pm 0.01 & 0.88\pm 0.01\\
B (2\ {\rm GeV}) [{\rm GeV}] & 2.30\pm 0.06 & 2.31\pm 0.06\\
F [{\rm MeV}]                & 86.5\pm 0.2 & 86.6\pm 0.3 \\
\bar\ell_3       & 2.7\pm 0.5 & 2.9\pm 0.6\\
\bar\ell_4     & 4.1\pm 0.2 & 4.0\pm 0.2 \\
\Sigma/\Sigma_0 & 2.11\pm 0.13 & 2.14\pm 0.16\\
B/B_0          & 1.28\pm 0.07 & 1.31\pm 0.10\\
F/F_0         & 1.29\pm 0.04 & 1.28\pm 0.04\\[0.15ex]
\hline
\phantom{xx}&&\\[-2.4ex]
f_+(0)      & 0.975\pm 0.006 & 0.975\pm 0.006\\
\Delta_{CT}\cdot 10^3  & 4.8\pm 5.7 & 3.8\pm 5.8\\
\Delta'_{CT}\cdot 10^3 & -70\pm 28   & -68\pm 29\\[0.15ex]
\hline
\phantom{xx}&&\\[-2.4ex]
\langle r^2\rangle_V^{K^+} [{\rm fm}^2]  & 0.224\pm 0.129 & 0.225\pm 0.129 \\
\langle r^2\rangle_V^{K^0} [{\rm fm}^2]  & -0.026\pm 0.098 & -0.026\pm 0.097\\[0.15ex]
\hline
\phantom{xx}&&\\[-2.4ex]
F_\pi^2
        & 0.53 + 0.57 - 0.10 
        & 0.54 + 0.56 - 0.10\\
F_K^2      & 0.37 + 0.70 - 0.07 
        & 0.37 + 0.70 - 0.07\\
F_\pi^2M_\pi^2
        & 0.43 + 0.68 - 0.11
        & 0.42 + 0.69 - 0.11\\
F_K^2M_K^2
        & 0.30 + 0.78 - 0.08
        & 0.30 + 0.78 - 0.08\\
F_\pi F_K f_+(0)
       & 0.45 + 0.66 - 0.11
       & 0.46 + 0.66 - 0.12 \\[0.15ex]
\hline
\phantom{xx}&&\\[-2.4ex]
\chi^2/N & 33.6/20 &  33.2/19\\[0.15ex]
\hline
\end{array}
\end{equation*}
\caption{Results of two different fits of the data from the RBC/UKQCD~\cite{Allton:2008pn,Boyle:2007qe,Boyle:2010bh} on pseudoscalar masses and decay constants, as well as on $K_{\ell 3}$ 
form factors. We considered  all the available unquenched data (Total), and either fixed the lattice strange quark mass (marked then with a star) or 
let it vary freely. Only statistical errors are shown and LECs are given at the scale $\mu =m_\rho$. The penultimate set of rows collects the relative fractions of LO/NLO/remainders at the minimum for
decay constants, masses and $K_{\ell 3}$ form factor at vanishing 
transfer momentum.} \label{table:UKQCDRBCtotal}
\end{table}
Our results are summarised in Tables~\ref{table:resultstotal} and  \ref{table:UKQCDRBCtotal}. The first series of rows corresponds to the outcome of the fit,  whereas the lower rows are quantities derived from the results of the fit (LO LECs, NLO LECs, quantities in the $N_f=2$ chiral limit, $K_{\ell 3}$ quantities, relative fraction of LO/NLO/remainders contributions at the minimum for several observables), and the last row is the $\chi^2$ per degree of freedom. Most of the comments made in Sec.~\ref{sec:discussion} can be restated, with a few changes in the case of the RBC/UKQCD data (larger value of $F_K/F_\pi$ and lower value of $f_+(0)$ than in the "Subset" case). We notice that the fits are fairly good, and that all NNLO remainders turn out to lie within their expected range.

\end{document}